\documentclass[a4paper,oneside,11pt]{scrartcl}

\usepackage[english]{babel} 				% Silbentrennung
\usepackage[T1]{fontenc}					% Zeichensatzkodierung
\usepackage[utf8]{inputenc} % Umlaute

\usepackage{times}

\usepackage{graphicx}	 
\usepackage{tabularx}

\usepackage{amsmath}
\usepackage{amssymb}
\usepackage{mathrsfs}   
\usepackage{braket}   	% Klammerngroesse anpassen
\usepackage{empheq}  	% mark a whole math display with a symbol es. to number each equation of a system
\usepackage{amsfonts}
\usepackage{amsthm}							% Umgebungen für Theoreme, Sätze,Definitionen...

\usepackage[authoryear,round]{natbib} %Zitationsstil

\usepackage{dsfont}
\usepackage{mathtools}

\def\XXint#1#2#3{{\setbox0=\hbox{$#1{#2#3}{\int}$} \vcenter{\hbox{$#2#3$}}\kern-.5\wd0}}

\newcommand{\R}{\mathbb{R}}

\newcommand{\vv}[1]{\boldsymbol{#1}}
\newcommand{\rot}{\nabla \times}

\renewcommand{\vec}{\boldsymbol}

\newcommand{\norm}[1]{\left\| #1\right\|}

\usepackage{color}    
\usepackage{marginnote}  %marginnotes always working
\usepackage{booktabs} % for much better looking tables
\usepackage{graphicx}   
\usepackage{subfig}    
\usepackage{footnote} %prints the footnotes in a minipage as a normal footnote
\usepackage{pdfpages} %allow to insert pages from other pdf files in the document
\usepackage{authblk} %author affiliation
\usepackage[unicode=true,bookmarks=true,bookmarksnumbered=false,bookmarksopen=false, breaklinks=false,pdfborder={0 0 0}, backref=false,colorlinks=false]{hyperref}   % internt links

\usepackage{xcolor}
\usepackage[colorinlistoftodos, textwidth=1.7in, textsize=footnotesize, shadow]{todonotes}

\newtheoremstyle{break}% name
{13pt}%      Space above, empty = 'usual value'
{13pt}%      Space below
{\itshape}% Body font
{}%         Indent amount (empty = no indent, \parindent = para indent)
{\bfseries}% Thm head font
{.}%        Punctuation after thm head
{5pt}% Space after thm head: \newline = linebreak
{}%         Thm head spec

\newtheoremstyle{definition}% name
{13pt}%      Space above, empty = 'usual value'
{13pt}%      Space below
{\normalfont}% Body font
{}%         Indent amount (empty = no indent, \parindent = para indent)
{\bfseries}% Thm head font
{.}%        Punctuation after thm head
{5pt}% Space after thm head: \newline = linebreak
{}%         Thm head spec

\newtheoremstyle{example}% name
{13pt}%      Space above, empty = 'usual value'
{13pt}%      Space below
{\normalfont}% Body font
{}%         Indent amount (empty = no indent, \parindent = para indent)
{\bfseries}% Thm head font
{.}%        Punctuation after thm head
{5pt}% Space after thm head: \newline = linebreak
{}%         Thm head spec

\theoremstyle{break}

\theoremstyle{break}

\theoremstyle{break}

\theoremstyle{definition}

\theoremstyle{example}

\allowdisplaybreaks[1] % allows page break within align, equations....

\begin{document}
\title{When Fields Are Not Degrees of Freedom}
\author{Vera Hartenstein}
\affil{LMU Munich, Mathematical Institute}
\author{Mario Hubert}
\affil{Columbia University, Department of Philosophy}

\date{June 13, 2018 \\
\vspace{1.5 em}
{\normalsize Forthcoming in  \emph{The British Journal for the Philosophy of Science
}}}

\maketitle
\begin{abstract}
We show that in the Maxwell--Lorentz theory of classical electrodynamics most initial values for fields and particles lead to an ill-defined dynamics, as they exhibit singularities or discontinuities along light-cones. This phenomenon suggests that the Maxwell equations and the Lorentz force law ought rather to be read as a system of delay differential equations, that is, differential equations that relate a function and its derivatives at \emph{different} times. This mathematical reformulation, however, leads to physical and philosophical consequences for the ontological status of the electromagnetic field. In particular, fields cannot be taken as independent degrees of freedom, which suggests that one should not add them to the ontology.
\end{abstract}

\thispagestyle{empty}

\newpage

\tableofcontents

\newpage

\section{The Ontology of Electromagnetism: Fields, Particles, or Both?}
\setcounter{page}{1}

What is the ontology of classical electrodynamics? The Maxwell--Lorentz theory, the most famous formulation of classical electrodynamics, mathematically introduces particles and fields. Particles obey the Lorentz force law, and fields obey the Maxwell equations. But it would be naive to reify all mathematical objects appearing in the formulation of a physical theory. And so we may debate three options for an ontology: a pure field ontology, a pure particle ontology, or a dualistic ontology comprised of particles and fields. A pure field ontology is no longer defended nowadays, although \citet{Mie:1912aa,Mie:1912ab,Mie:1913aa} and \citet{Weyl:1921aa} searched in this direction \citep[see][]{Smeenk:2007aa} and also Faraday had this idea (see \citealp[][]{Heimann:1971aa}; \citealp[][Ch.\ 6]{Lange:2002ys}). The general reading of the Maxwell--Lorentz theory indeed is that it poses the existence of fields and point particles alike. The dynamics is such that fields act on particles, and particles act on fields.

This dualistic ontology of point-like particles plus fields results in the well-known self-interaction problem: The Maxwell--Lorentz theory doesn't provide a law of motion for a particle affected by its \emph{own} field. The problem arises because the Lorentz force of the self-field on the charge is undefined at the position of the charge. 

In a recent paper, \citet{Lazarovici:2016aa} used the self-interaction problem as his core argument to emphasize the shortcomings of fields and to defend the action-at-a-distance theory advocated by Wheeler and Feynman. In this paper, we present an additional argument  \citep[briefly mentioned in][]{Lazarovici:2016aa}, independent of the self-interaction problem, that we use against the existence of fields while holding on to the Maxwell-Lorentz formulation of electrodynamics. The argument is based on the recent physical results of \citet{Deckert:2016aa}.

We will demonstrate what they did in detail and discuss in depth implications for the ontology of fields. Whereas we don't adhere to any specific theory to replace the Maxwell--Lorentz theory, the main claim is that the degrees of freedom of the electromagnetic field can be entirely reduced to the degrees of freedom of particles. Therefore, we argue that the Maxwell--Lorentz theory is rather to be interpreted as an action-at-a-distance theory disguised as a field theory. Before presenting the argument, we briefly introduce the Maxwell--Lorentz theory, as far as necessary, thereby outlining in passing its major mathematical problems.

\section{The Maxwell--Lorentz Theory}
\label{sec:intro}

In our jargon, the Maxwell--Lorentz theory uses the matter model of point charges, as opposed to smeared-out charges, which, often, is referred to as the Abraham model.
The charge density of a point-charge at space time point $(t, \vv x)$ is given by $\varrho_t(\vec{x})=\delta(\vec{x}-\vec{q}_t)$, where the Dirac $\delta$-distribution encodes that the charge is point-like and hence only concentrated on its trajectory $t \mapsto \vec{q}_t$, 
and we choose units where the speed of light and the charge are both $1$. The corresponding density current is 
the charge density multiplied by the velocity of the charge, namely (in standard non-relativistic notation), $\vec{j}_t(\vec{x})=\varrho_t(\vec{x})\vec{v}_t=\delta(\vec{x}-\vec{q}_t)\vec{v}_t$. Then, the time evolution of $N$ charged particles and their fields is given by the coupled system of the Lorentz equations
\begin{equation}
\label{eq:lorentz}
\frac{d}{dt} \begin{pmatrix}
\vv q_{i,t} \\ \vv p_{i,t}
\end{pmatrix} = \begin{pmatrix}
\vv v(\vv p_{i,t}) \\ \sum_{j=1}^N \vv E_{j,t}(\vv q_{i,t}) + \vv v(\vv p_{i,t}) \times \vv B_{j,t}(\vv q_{i,t}) 
\end{pmatrix}
\end{equation}
and Maxwell equations
\begin{equation}
\label{eq:maxwell}
\frac{d}{dt} \begin{pmatrix}
 \vv E_{i,t} \\ \vv B_{i,t}
\end{pmatrix} = \begin{pmatrix}
 \rot \vv B_{i,t} - 4 \pi \vv v ( \vv p_{i,t}) \delta ( \cdot - \vv q_{i,t}) \\ - \rot \vv E_{i,t}
\end{pmatrix}
\end{equation}
\begin{equation}
\label{eq:maxwell_constraints_particle}
\nabla\cdot\vec{E}_{i,t} =4\pi \delta(\vec{x}-\vec{q}_{i,t}),	 \qquad \nabla\cdot\vec{B}_{i,t}=0,
\end{equation}
for all charges $i\in\{1,\ldots, N\}$. Here, $\vv q_{i,t}$ and $\vv p_{i,t}$ denote the position and the momentum of the $i$th particle; $\vv v(\vv p_{i,t})=\frac{\vv p_{i,t}}{\sqrt{{\vv p_{i,t}}^2+m^2}}$ is the relativistic velocity defined by the $i$th momentum, where we assume all particles to have the same mass $m$; and $\vv E_{i,t}$ and  $\vv B_{i,t}$ are the electric and magnetic fields associated with the $i$th particle at time $t$.

In this representation, we follow \cite{Deckert:2016aa}, who split the total electromagnetic field into $N$ sub-fields corresponding to single particles, which is unproblematic due to the linearity of the Maxwell equations. Moreover, $\vv E_{i,t}$ and  $\vv B_{i,t}$ may contain at this stage free-field parts; in the course of the paper, however, we show how to get rid of those free fields.

Special solutions of the field equations \eqref{eq:maxwell} and \eqref{eq:maxwell_constraints_particle} for single predetermined charge trajectories, the advanced and retarded \emph{Liénard–Wiechert} fields, are known. The electric field component of the advanced (+) and retarded (-) Liénard–Wiechert field of a single particle is given by
\begin{equation} \label{eq:lw_e_field}
\vec{E}^\pm_t(\vec{x})=\frac{1}{4\pi}
\left(
\underbrace{\frac{(1-\vec{v}^2)(\vec{n}\pm\vec{v})}{(1\pm\vec{v}\cdot\vec{n})^3\lvert \vec{x} - \vec{q}\rvert^2}}_{\text{near field}}+
\underbrace{\frac{\vec{n}\times\left((\vec{n}\pm\vec{v})\times\dot{\vec{v}}\right)}{\left(1\pm\vec{v}\cdot\vec{n}\right)^3 \lvert \vec{x} - \vec{q}\rvert}}_{\text{radiation field}}
\right),
\end{equation}
and the magnetic field can be calculated from
\begin{equation} \label{eq:lw_b_field}
\vec{B}^\pm_t(\vec{x})=\mp\vec{n}\times\vec{E}^\pm_t(\vec{x}).
\end{equation}
\citep[see for instance][Section 2.1]{Spohn:2004aa}. We see that the Liénard--Wiechert field has two parts: the near field and the radiation field. The near field is always attached to the particle and descends like $\frac{1}{\vv x^2}$; it is the dominating part in the vicinity of the particle and negligible far away from it. The radiation field, in contrast, depends on the acceleration of the particle, and it is the dominating part far away, for it descends like $\frac{1}{\vv x}$. In addition, it also increases near the particle's position.

\begin{figure}[ht]
\centering
\includegraphics[width=7cm]{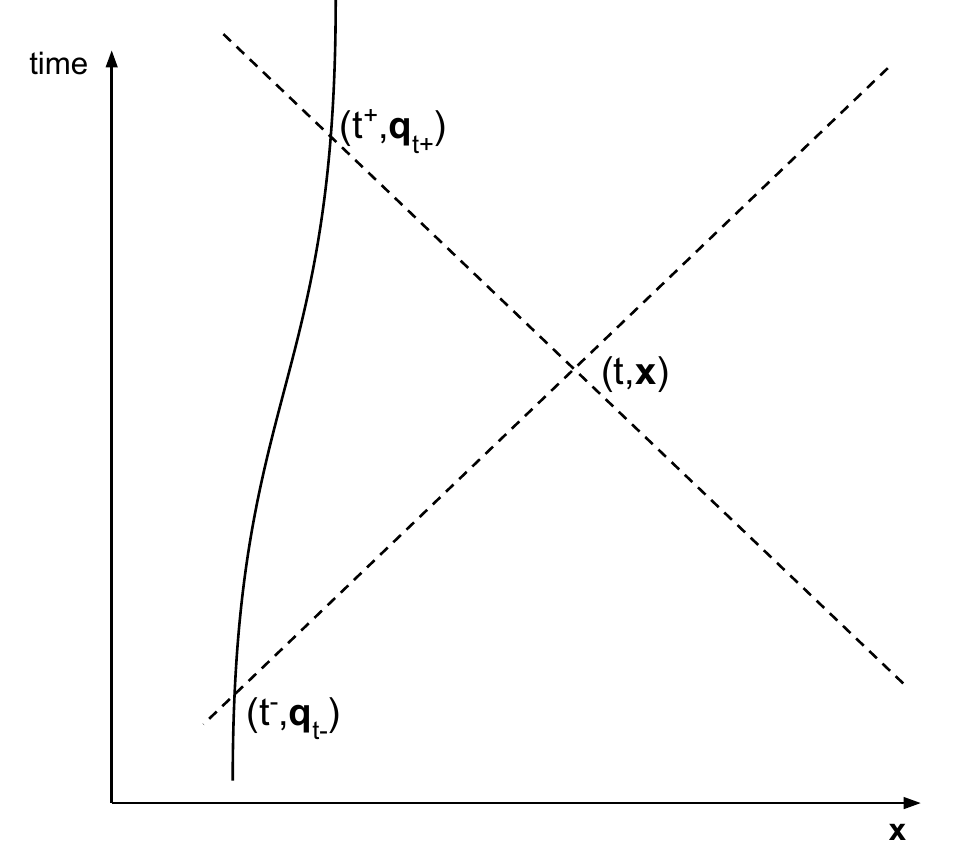}
\caption{Depiction of retarded times $t^-$ and advanced times $t^+$ relative to $t$ in a space-time diagram. The dashed lines represent the light-cone at $(t,\vec{x})$, which is crossed twice by a trajectory $t^\prime \mapsto \vec{q}_{t^{\prime}}$.}
\label{fig:particle_trajectory}
\end{figure}

For the retarded Li\'enard-Wiechert field, on the right side of (\ref{eq:lw_e_field}), the quantities $\vec{q}$, $\vec{v}$, $\dot{\vec{v}}$, and $\vec{n}$ have to be evaluated at the retarded time $t^-$, which is implicitly defined by 
$
t^-=t-\lvert \vv x - \vec{q}_{t^-}\rvert.
$ It is the time at which the backward light-cone with apex at $(t,\vec{x})$ crosses the world-line  $t^{\prime} \mapsto \vec{q}_{t^{\prime}}$ of the particle (see Fig.\ \ref{fig:particle_trajectory}). The vector $\vec{n}$ is a spatial unit vector derived from the position of the particle, namely,
$
\vec{n}^{-}:=\frac{\vec{x}-\vec{q}_{t^-}}{\lvert\vec{x}-\vec{q}_{t^-}\rvert}.
$

The advanced fields are evaluated at the advanced time $t^+$ given by
$
t^+=t+\lvert \vv x - \vec{q}_{t^+}\rvert.
$
The advanced time lies on the \emph{forward} light-cone with apex $(t,\vec{x})$ intersecting the world-line of the particle (see Fig.\ \ref{fig:particle_trajectory}). In analogy to the retarded case, we can define a spatial unit vector $\vec{n}^{+}$. Advanced solutions are often abandoned in application since accepting them would amount to backward causation. This phenomenon seems to contradict our experience and is not confirmed by experiments, so that only the retarded fields are meant to be physically significant \citep[see, for instance,][Ch.\ 3]{Price:1996fk}. 

As the Maxwell equations \eqref{eq:maxwell} and \eqref{eq:maxwell_constraints_particle} are linear, any solution $t \mapsto \vec{f}_t=(\vec{E}_t,\vec{B}_t)$ can be written as a convex combination of the retarded and advanced Li\'enard--Wiechert fields, $\vec{f}_t^-$ and  $\vec{f}_t^+$, plus some solution $\vec{f}_t^{\text{free}}$ of the charge-free Maxwell equations:
\begin{equation}
\label{eq:6}
\vec{f}_t=\lambda \vec{f}_t^- + (1-\lambda)\vec{f}_t^+ + \vec{f}_t^{\text{free}},
\end{equation}
with $\lambda \in [0,1]$.

In order to calculate the Li\'enard--Wiechert fields \eqref{eq:lw_e_field} and \eqref{eq:lw_b_field}, one presupposes the trajectory of the particle. But if both fields and trajectories are  unknown, we need to couple the Maxwell equations \eqref{eq:maxwell} with the Lorentz equations \eqref{eq:lorentz} and compute fields and trajectories simultaneously. 
In order to compute the Lorentz force one needs to evaluate the self-field at the particle's position, and this procedure is undefined as \emph{both} denominators of \eqref{eq:lw_e_field} become zero. So the Maxwell--Lorentz theory fails to deliver a solution---even for the simplest physical system consisting of one moving charge in its own field. 

In essence, there are two broad strategies to cope with this problem. One is the Abraham model, where particles are tiny balls with non-zero diameter \citep[see][]{Abraham:1908aa,Lorentz:1916aa,Spohn:2004aa}. The other strategy is to remedy the field without touching the size of the particles. The obvious way would be to adjust the Maxwell equations such that they no longer lead to the self-interaction problem. The Bopp--Podolsky theory does it with linear, but higher-order, field equations, while the Born--Infeld theory has non-linear field equations (see Appendix \ref{sec:appendix}). 
Or one could modify the Lorentz equations \eqref{eq:lorentz} and replace the summand $j=i$ by a well-defined self-interaction term, like for instance the Lorentz--Abraham--Dirac term \citep[cf.][]{Dirac:1938aa}, the Landau--Lifshitz term \cite[cf.][]{Spohn:2004aa}, or as a starting point for a study also just zero.
A more radical way is to get rid of fields (and thus, of the ill-defined self-interaction) in the first place and construct an action-at-a-distance theory (see Appendix \ref{sec:wheeler-feynman}).\footnote{For a detailed philosophical discussion of how to deal with the self-interaction problem we refer to \citet{Frisch:2005fr}.}

In the standard literature on the Maxwell--Lorentz theory, one usually encounters two scenarios: 
Either the trajectories of particles are given and the Maxwell equations are solved, or fields are given and the trajectories are calculated. In both cases, one gets accurate solutions. The fully coupled problem is usually ignored \citep[see][p.\ 745]{Jackson:1999aa}. Requiring accurate solutions for the coupled system is not mathematical pedantry; on the contrary, the self-field accounts for radiation damping, the effect that a radiating particle has to change its motion because it loses energy. And this change in motion due to self-interaction is (although small) a crucial measurable physical effect! Since the effects of the self-field are not relevant for most practical purposes, the Maxwell--Lorentz theory makes mostly successful empirical predictions.

\begin{figure}[ht]
    \begin{center}
       \includegraphics[width=0.6\linewidth]{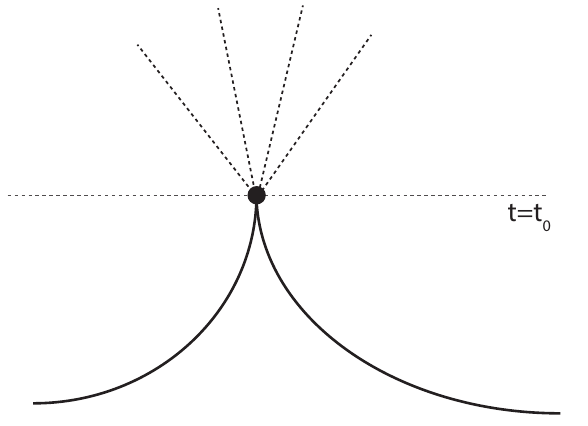} 
    \end{center}
    \caption{Two particles collide at $t=t_{0}$. Due to the singular factor $\frac{1}{|\vec{x}_i-\vec{x}_j|^2}$ in the force law, there is no unique dynamics after collision, which is indicated by the black dashed lines.}
    \label{fig:gravity}
\end{figure}

A second singularity of the theory corresponds to the $N$-body problem of Newtonian gravitation: the collision of particles is not well-defined \citep[see, for instance,][]{Heggie:2006aa}. Since the gravitational force goes like $\frac{1}{|\vec{x}_i-\vec{x}_j|^2}$, the dynamics breaks down when particles collide, that is, when $\vec{x}_i(t)\xrightarrow[t\rightarrow t_0]{} \vec{x}_j(t_0)$ for some finite time $t_0$ (see Fig.\ \ref{fig:gravity}). There is hence no further time evolution after collision. It may be possible to extend the trajectories after collision, but this extension is not unique. Although there may be collisions when $N$ particles move in the gravitational field, the initial conditions resulting in these collisions have measure zero \citep{Saari:1973aa}.
We encounter the same problem for charged particles. The Liénard--Wiechert fields \eqref{eq:lw_e_field} contain in the near field the factor $\frac{1}{|\vv x-\vv q|^2}$, which blows up when particles are about to collide. In order to have well-defined dynamics one needs to make sure that particles cannot come arbitrarily close to each other. If they still do, further equations would be needed to describe the future motion. But we would expect, as has been rigorously shown for classical mechanics, that initial configurations leading to collisions are atypical, that is,  have measure $0$. Then, one could ignore for all practical purposes dynamics leading to collisions. 

\begin{figure}[ht]
    \begin{center}
       \includegraphics[width=0.6\linewidth]{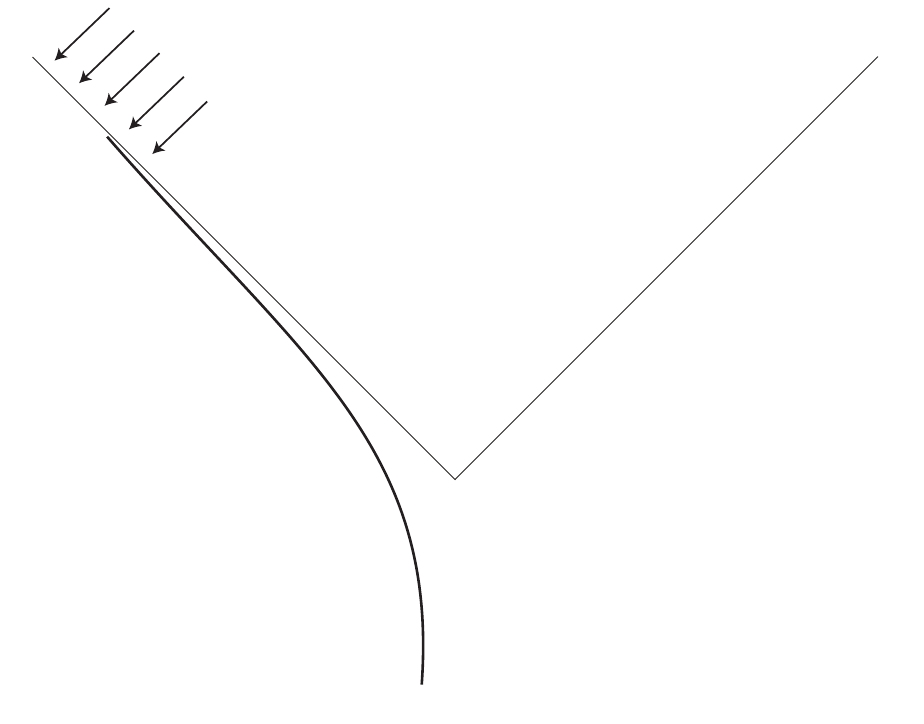} 
    \end{center}
    \caption{A particle approaching the speed of light. In this case the electromagnetic field would accumulate on the light-cone (marked by the arrows).}
    \label{fig:runaway}
\end{figure}

A third singularity arises in the term $\frac{1}{(1 \pm \vv n \cdot \vv v)^3}$ of the Liénard--Wiechert fields. \citet{Dirac:1938aa} found out that the theory allows for runaway solutions, that is, solutions that approach the speed of light exponentially fast (see Fig.\ \ref{fig:runaway}). When particles do so, this term approaches infinity. There are two problems with this kind of solutions. First, we do not observe such accelerating particles. Second, such a particle needs to constantly radiate, and this very radiation would accumulate on the light-cone leading to high-energy radiation (see also Fig.\ \ref{fig:runaway}). Such a phenomenon is not observed, either. 

\section{The Problem of Initial Values}
\label{sec:problem-iv}
We now turn to the problem of initial values as mathematically presented by \citet{Deckert:2016aa}. 
Solving the coupled system of Maxwell's and Lorentz's equations \eqref{eq:lorentz}-\eqref{eq:maxwell_constraints_particle} for point charges \emph{without} self-interaction reveals a mathematical fact: most initial conditions lead to singularities or discontinuities on future light-cones---we call these pathologies on light-cones \emph{shock fronts}. 
As we shall explain, this observation questions the initial-value formulation of the theory and finally the ontological status of fields. 

\subsection{The Existence of Shock Fronts}
\label{sec:ME}

Let's start with an example.
Consider a single charged particle with given trajectory $t \mapsto \vv q_t$ and a predetermined initial field $\vv f_0$ on a space-like hyper-surface, say $\lbrace t=0\rbrace$. This allows us to compute the electromagnetic fields of that particle at any space-time point.
The field is composed of the time-evolved initial field and the radiated field of the particle. 

\begin{figure}[ht]
    \begin{center}
       \includegraphics[width=0.8\linewidth]{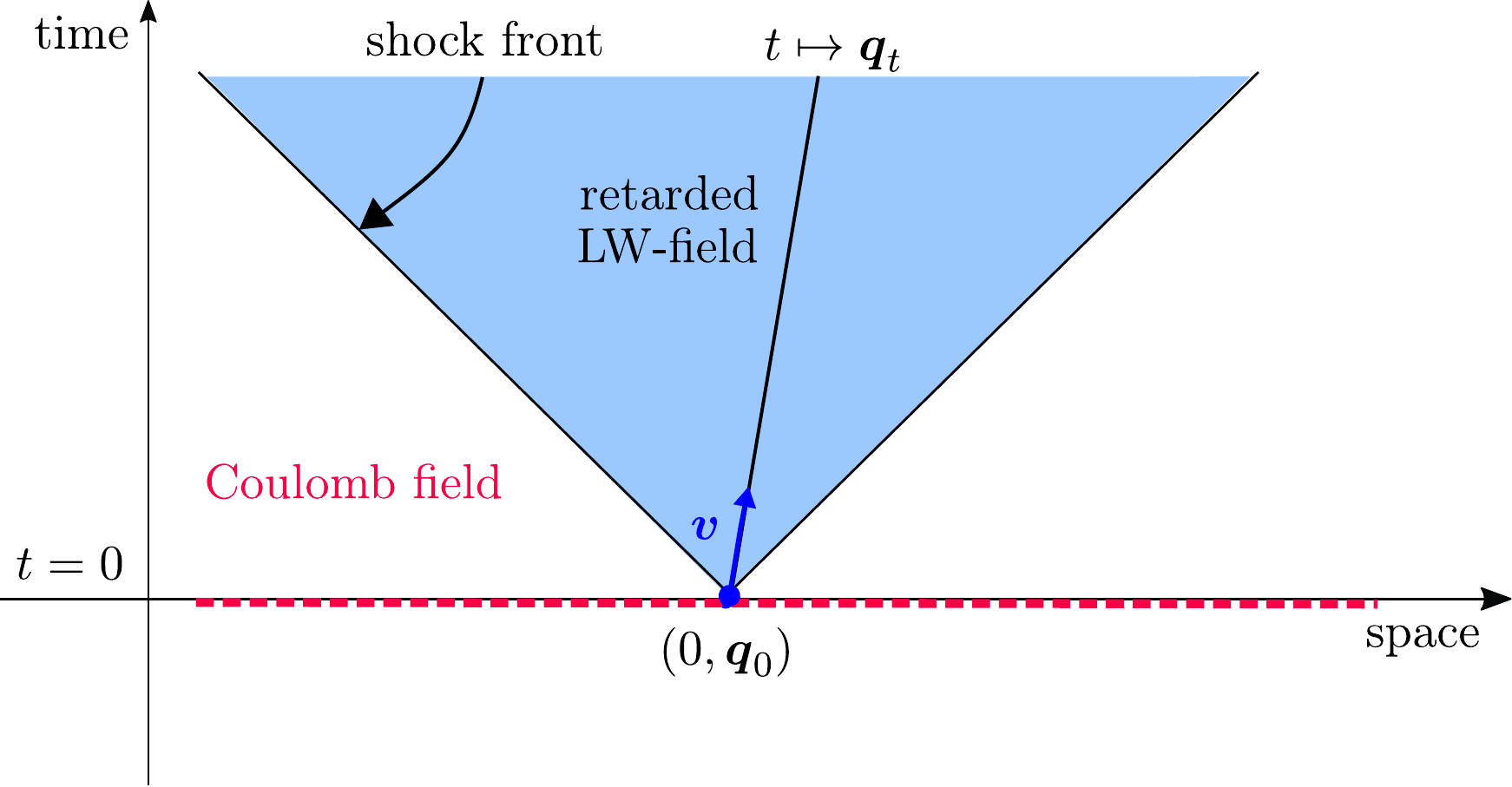} 
    \end{center}
    \caption{Illustration of supports of the (one-particle) field components.}
    \label{fig:coulomb-field}
\end{figure}

Say the particle moves with constant velocity $\vv v$ and the initial field $\vv f_0$ is just the Coulomb field
\begin{equation}
\label{eq:coulomb}
\vv f_0(\vv x) = \begin{pmatrix}
\frac{\vv x-\vv q_0}{|\vv x - \vv q_0|^3} \\ 0
\end{pmatrix}.
\end{equation}
Then, one finds the field components as illustrated in Fig.\ \ref{fig:coulomb-field}.
Inside the future light-cone of $(0,\vv q_0)$ a field component due to the 
charge trajectory $t\mapsto \vv q_t$ builds up.
The initial Coulomb field $\vv f_0$ remains outside the future light-cone of $(0,\vv q_0)$; that's not surprising as the field freely evolves in this regime.
In addition, there is a distribution that only depends on the initial position and momentum $\vv q_0, \vv p_0$ and which has support on the boundary of the light-cone---even though the initial Coulomb field $\vv f_0$ is smooth.

Are these shock fronts supposed to be there, or has something gone wrong with the choice of the initial configuration?
The only mathematical condition on initial values that the Maxwell--Lorentz theory dictates are the Maxwell constraints \eqref{eq:maxwell_constraints_particle} (at time $t=0$), which the Coulomb field complies with.
Though the constructed toy problem matches all requirements of the Maxwell--Lorentz theory, it exhibits question-begging behavior of the field.

In fact, the exact field equation in \citet{Deckert:2016aa} reveals that one can get rid of this singularity when we change the initial conditions of the charge to $\vv p_0 = \vv v = 0$. 
Then, and only then, the delta distribution located on the light cone vanishes.  This, however, shows that the initial velocity is no longer a free variable; rather, it has to be consistent with the initial field.

If the velocity $\vv v$ is chosen equal to zero, there is yet another feature in the field: If the initial acceleration of the charge is non-zero, the field shows a discontinuous jump on the light-cone. Here is why. For continuity,
the retarded Li\'enard-Wiechert field $\vv f_t^-$ has to match 
the Coulomb field $\vv f_0$ on the light-cone.  
There the Li\'enard-Wiechert field is a function of $\vv q_0, \vv v_0$, and $\dot{\vv v}_0$ since in this case the retarded time is $t^-=0$.
This implies that unless the initial acceleration is zero ($\dot{\vv v}_0=0$), $\vv f_t^-$ will differ from $\vv f_0$ on the future light-cone. 
So aiming at continuous fields not only the initial momentum but even the initial acceleration of the charge cannot be freely chosen.

\begin{figure}[ht]
    \begin{center}
       \includegraphics[width=0.8\linewidth]{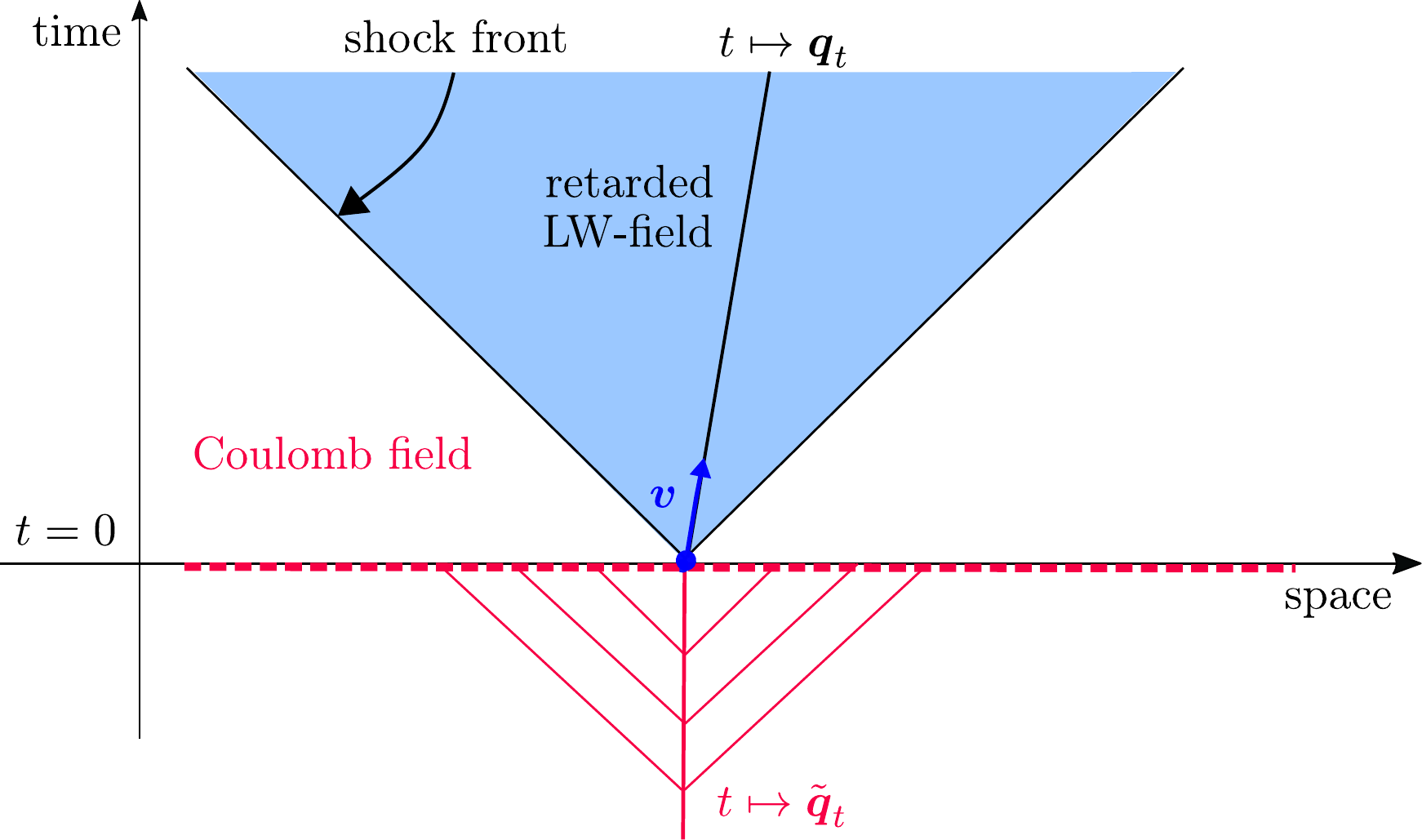} 
    \end{center}
    \caption{The auxiliary trajectory $t \mapsto\tilde{\vv q}_t$ for $t \leq 0$ attaches with a kink to $t \mapsto \vv q_t$ for $t\geq 0$ at space-time point $(0,\vv q_0)$.}
    \label{fig:coulomb2}
\end{figure}

At first sight, this phenomenon may seem surprising, but it has a rather
simple explanation. Each inhomogeneous field encodes the history of an auxiliary charge trajectory $t\mapsto \tilde{\vv q}_t$, 
a charge that has generated the field sometime somewhere in the past. In our case, the initial Coulomb field corresponds to the field generated by a charge that has been resting at position $\vv q_0$ during its entire past
(see Fig.\ \ref{fig:coulomb2}). 

If the auxiliary charge trajectory, or in other words the charge history, does not fit the future trajectory at time $t=0$ there will be a kink, and this sudden change of velocity and acceleration will result in a radiation field traveling along the future light-cone of $(0,\vv q_0)$. 
In other words, if $\vv v_0$ happens to be non-zero, an infinite acceleration is necessary to change it from $\tilde{\vv v}_0=0$ to $\vv v_0$, and the corresponding radiation gives rise to distributions,
whereas a step in the acceleration merely causes a discontinuity on the
light-cone.

\subsection*{The General Case}

The phenomenon of shock fronts can be extended to arbitrary initial fields and an arbitrary predetermined particle trajectory $t \mapsto \vv q_t$. 
Therefore, it is convenient to parameterize the initial field by means of 
a history (or auxiliary trajectory) $t \mapsto \tilde{\vv q}_t$ and an initial free field 
$\vv f_{0}^{\text{free}}$ (cf.\ \eqref{eq:6})
\begin{equation}
    \label{eq:f_0}
    \vv f_{0} =  \lambda \tilde{\vv f}_{0}^-+
    (1-\lambda) \tilde{\vv f}_{0}^+ + \vv
    f_{0}^{\text{free}},
\end{equation}
where the tilde over the advanced and retarded Li\'enard-Wiechert fields denotes, that these are functionals of $t \mapsto \tilde{\vv q}_t$.
This parameterization of the initial field is not unique and merely a choice of the free field. Nevertheless, it is sufficient in order to explain the problem with initial values. 

\begin{figure}[ht]
    \begin{center}
       \includegraphics[width=0.8\linewidth]{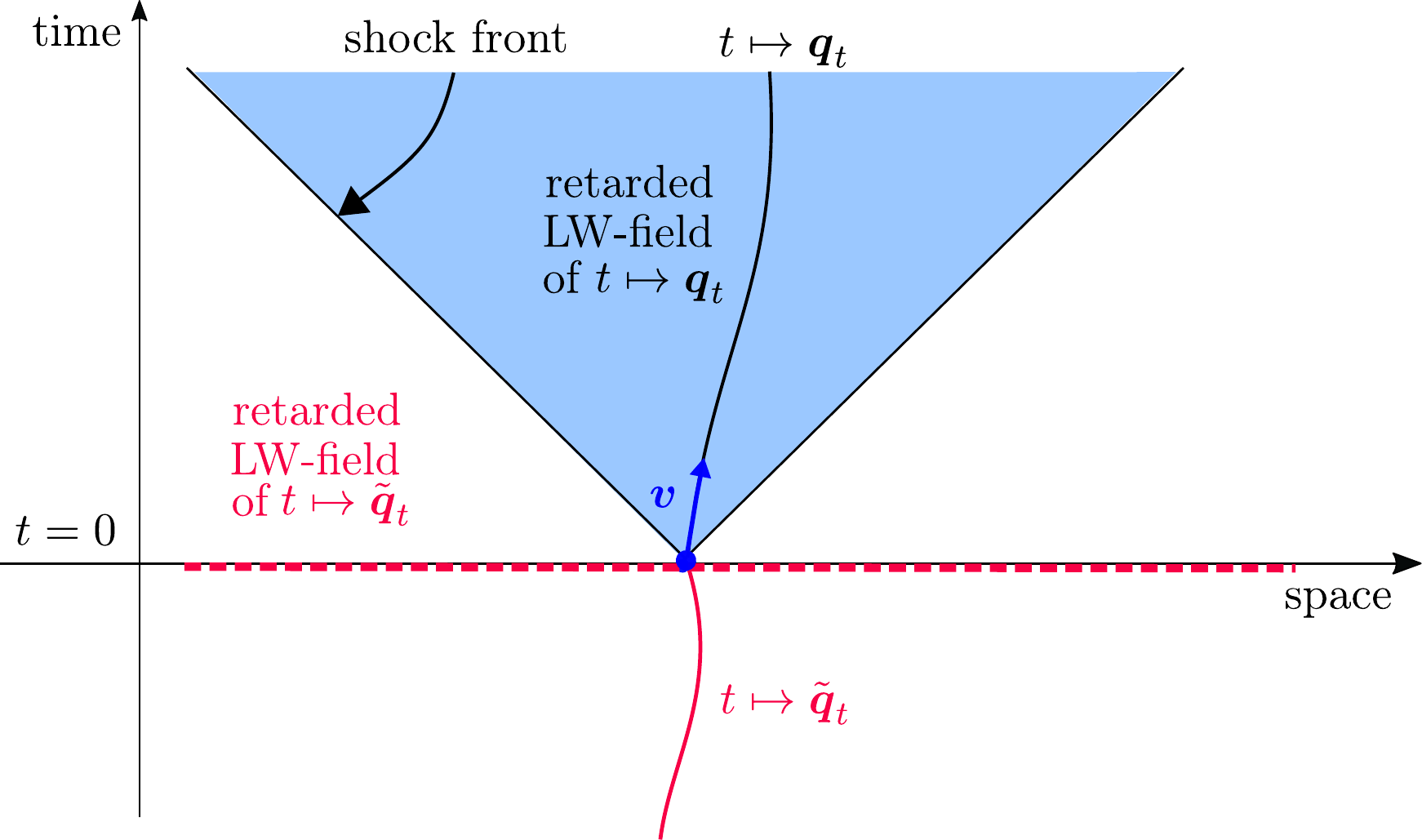} 
    \end{center}
    \caption{Supports of the single components of the general Maxwell field $\vec{f}_t$ of a one-particle system for the initial field $\vv f_0$. The initial free field $\vv f_{0}^{\text{free}}$ evolves independently of the particle and is supported all over the space-time.}
    \label{fig:general}
\end{figure}

The time-evolved field shows the same features as in the special case, as illustrated in Fig.\ \ref{fig:general}, namely, 
a distribution and discontinuity located along the light-cone boundary of $(0,\vv q_0)$ for generic initial fields.

The existence of discontinuities and singularities on the light-cones are a mathematical fact. In the one-particle case they don't pose any further problems because the particle is not affected by them. Once the motion of many particles is considered, the regularity of the electromagnetic field on the light-cones becomes crucial.

\subsection{How Shock Fronts Affect the Dynamics in Many-Particle Systems}

\label{sec:ML}

\begin{figure}[ht]
    \begin{center}
       \includegraphics[width=0.7\linewidth]{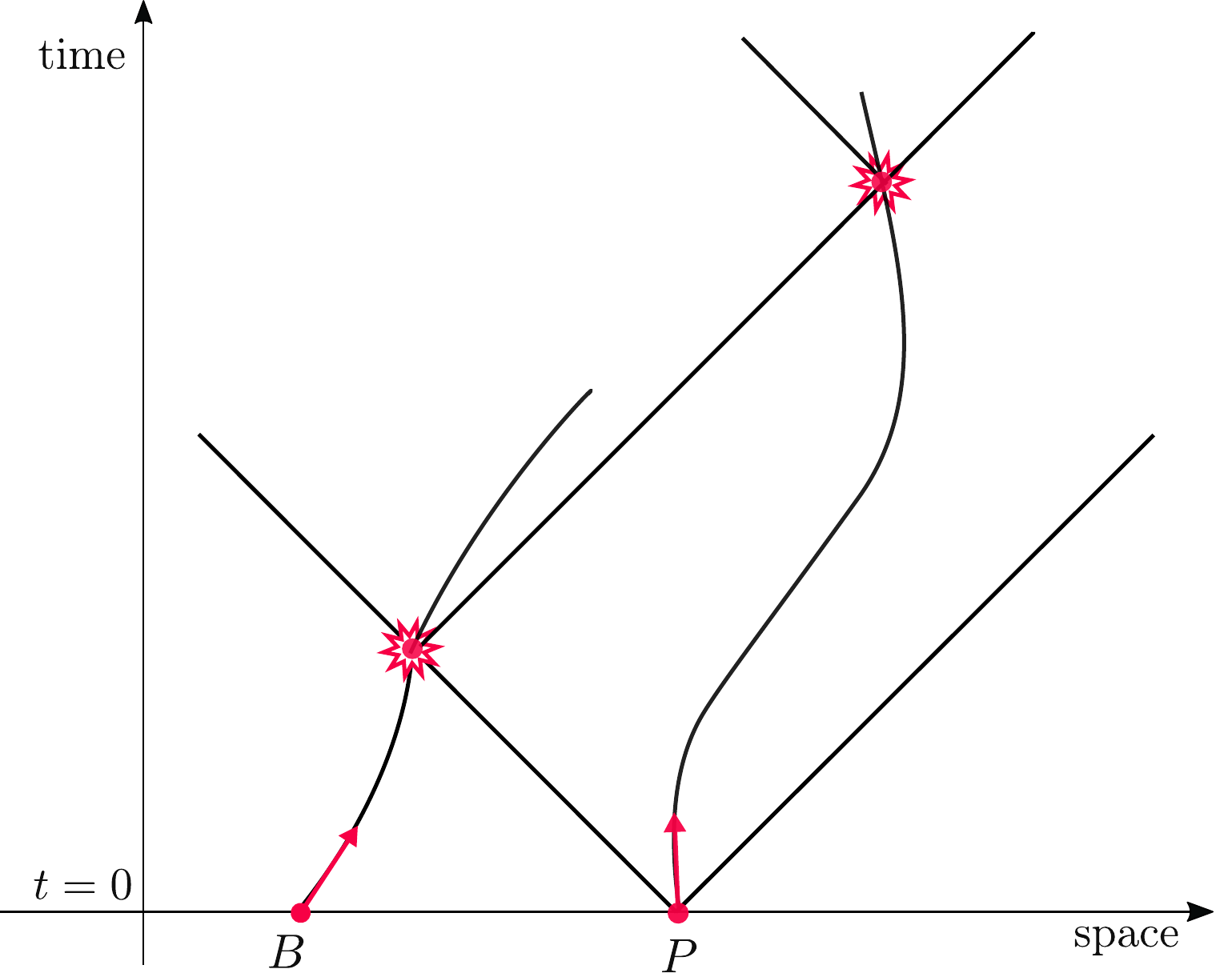} 
    \end{center}
    \caption{Illustration of a network of shocks. The left particle $B$ crosses the light-cone of the right particle $P$. $B$'s motion is affected by the pathologies on the light-cone. If $B$ hits a singularity, it's motion will end; if it hits a discontinuity, it will feel a kick (infinite acceleration). These kicks cause shocks on its own light-cone. If $P$ hits $B$'s light-cone, it will feel the shock.}
    \label{fig:network}
\end{figure}

Let's consider two charged particles, $P$ and $B$. If $B$ crosses $P$'s light-cone, $B$'s motion changes with respect to what happens on the light-cone (see Fig.\ \ref{fig:network}). Depending on the acceleration of $P$, we discussed three scenarios on the light-cone:
\begin{enumerate}
\item
The field on the light-cone is continuous.
\item
The field on the light-cone has a finite discontinuous jump.
\item
The field on the light-cone has a singularity ($\delta$-distribution).
\end{enumerate}

In the first case, nothing pathological happens. 
The other cases are more interesting. 
If $P$'s field jumps discontinuously, $B$ feels a kick, and this acceleration causes shock fronts on $B$'s light-cone. If  $P$, as in Fig.\ \ref{fig:network}, crosses this light-cone, it will also feel a kick that is transmitted on the light-cone, and so on. 

In the third case, when $B$ hits a singularity, it seems that the dynamics breaks down because the force acting on $B$ given by the Lorentz equation \eqref{eq:lorentz} requires the evaluation of the field due to particle $P$ on 
the light-cone, where it exhibits a delta distribution.
Thus, as in the case of colliding particles in Newtonian gravitation (see Sec.\ \ref{sec:intro}), there is no unique extension of the trajectory and the dynamics ends here. 

The more particles there are the more shocks there will be. If we imagine a more realistic system of $10^{13}$ particles, there will be a dense network of shocks that we ought to observe all the time. It turns out that the radiation created by a charge running into a shock front is quite strong, namely, of the order of $1$ Watt \citep[][p.\ 13]{Deckert:2016aa}. But we don't see such radiation! And so we need to restrict the initial values to avoid the creation of these shocks.

\subsubsection*{Bad Solutions Are Dense}

Now let's see how robust good global Maxwell--Lorentz solutions are. 
By ``good'' we mean that there are no shock fronts on light-cones (``bad'' would indicate shock fronts).
Therefore, assume we have an initial value
$(\vv q_{i,0}, \vv p_{i,0}, \vv f_{i,0})$
 that leads to a smooth global solution, which is a smooth function of time $t \mapsto (\vv q_{i,t}, \vv p_{i,t}, \vv f_{i,t}), t \in \R$. 
Then, in any arbitrarily small neighborhood of this initial datum, there are initial values $(\vv q'_{i,0}, \vv p'_{i,0}, \vv f'_{i,0})$ that generate shock fronts preventing the system to have a global solution. 
The dynamics breaks down at the time where the first charge runs into such a front.

\begin{figure}[ht]
    \begin{center}
       \includegraphics[width=0.7\linewidth]{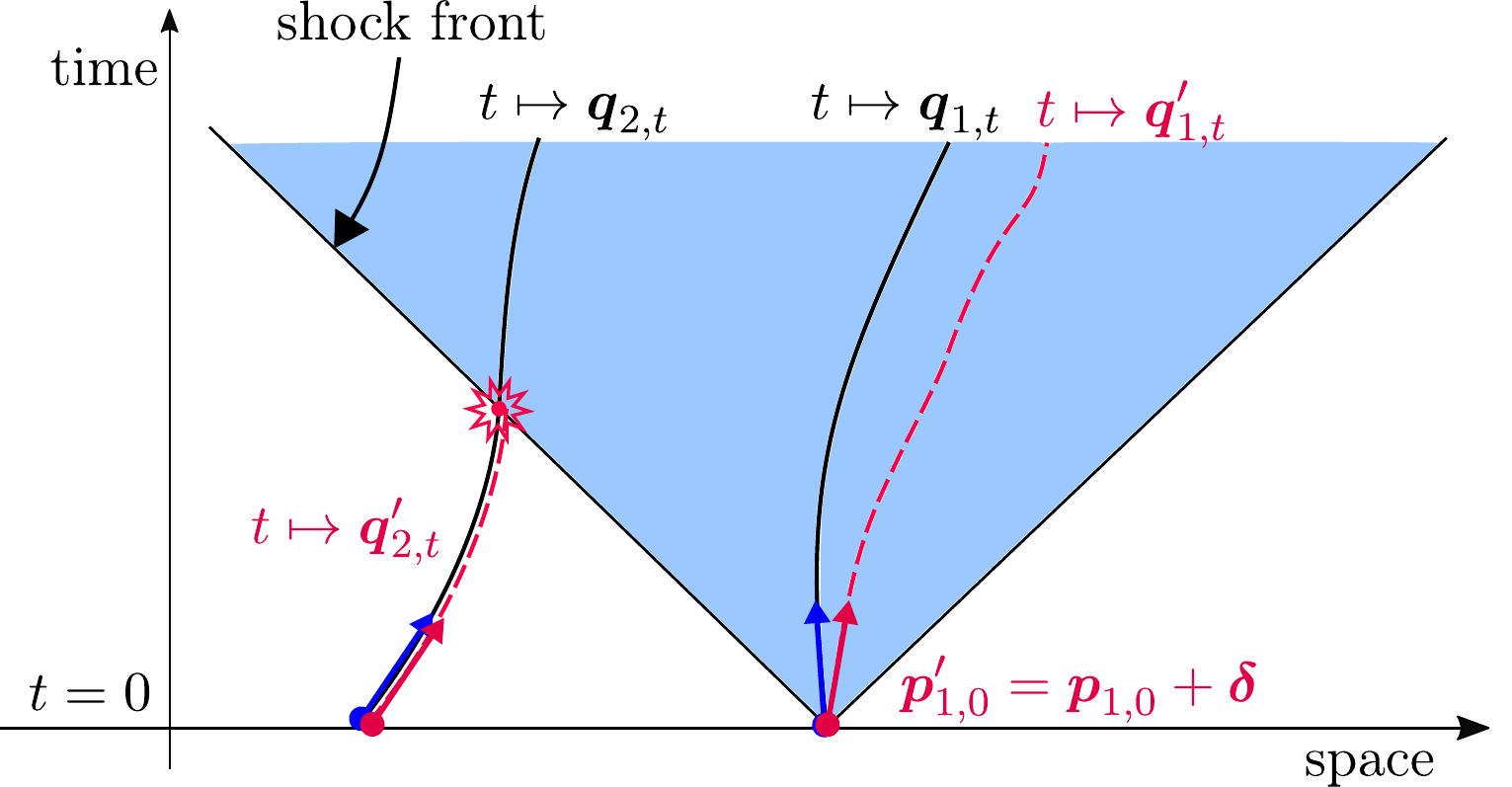} 
    \end{center}
    \caption{Perturbation of the initial momentum of charge $1$ at $t=0$. Red dashed trajectories are modified black trajectories. Within every $\vv \delta$ of the initial momentum of charge $1$ there is a trajectory that causes a singular shock front on the light-cone, which affects charge $2$ situated on the left.}
    \label{fig:perturbation-1}
\end{figure}

As suggested by \citet{Deckert:2016aa}, we briefly sketch two ways how to construct bad initial values.
First, one could transform the \emph{initial momentum} of, say, particle 1 by some $\vv \delta \neq 0$, which would create a distribution on the light-cone boundary, see Fig.\ \ref{fig:perturbation-1}. When particle 2 runs into this shock the dynamics stops.

\begin{figure}[ht]
    \begin{center}
       \includegraphics[width=0.7\linewidth]{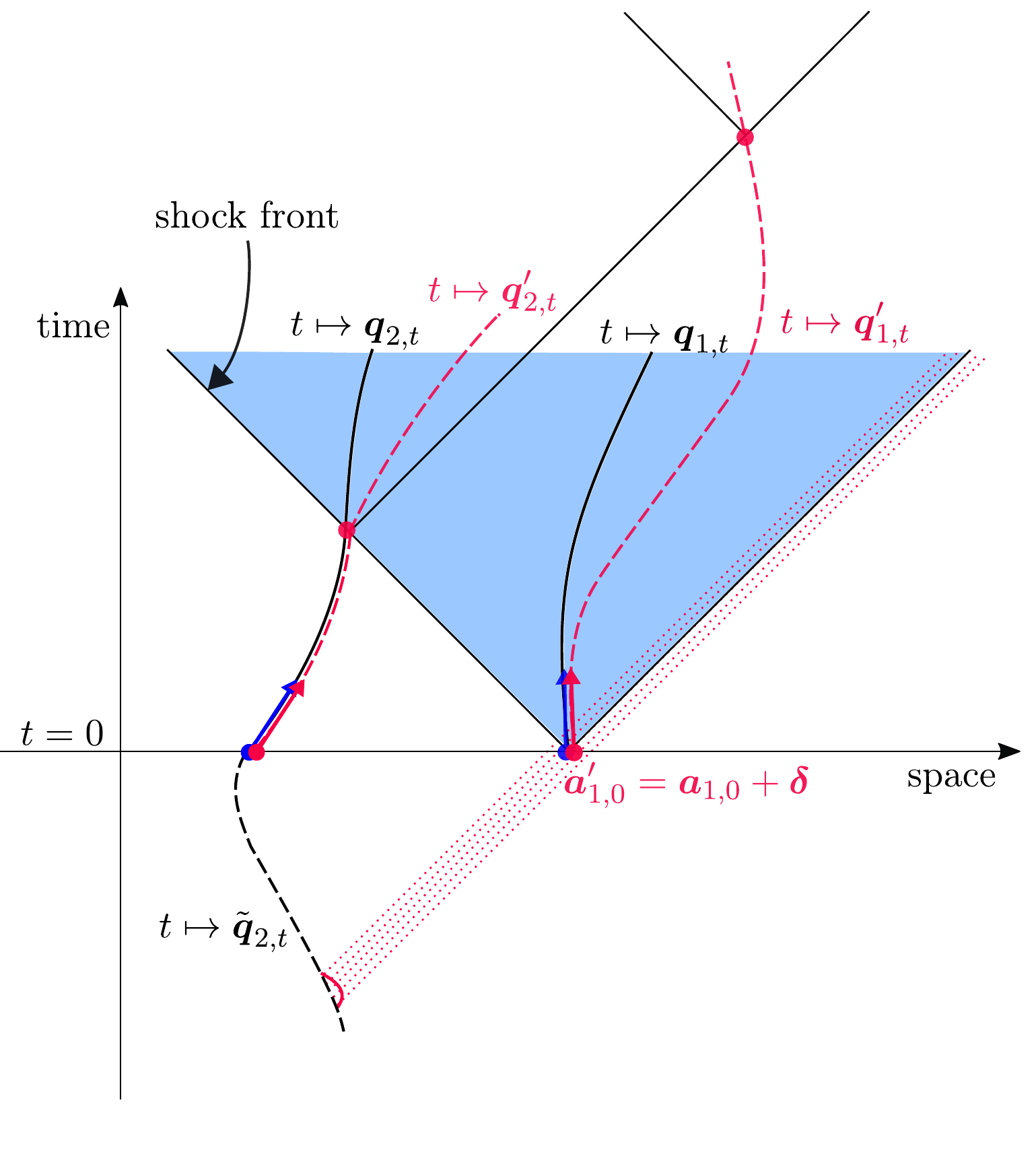} 
    \end{center}
    \caption{Perturbation of the trajectory of the left particle at some retarded time may change the field at $t=0$ and cause a discontinuous acceleration in the right particle, which leads to discontinuities on the light-cone. This may be the starting point of a future network of shocks.}
    \label{fig:perturbation-2}
\end{figure}

Second, one could slightly change the \emph{initial field} generated by, say, charge $2$ in the vicinity of the initial position of charge $1$ (see Fig.\ \ref{fig:perturbation-2}).
What happens in this case is a change of the Lorentz force on charge $1$ at time $0
$ so that the initial acceleration changes, which results in a discontinuity on the light cone of $(0, \vv q'_{1,0})$.
This time the shock is only a discontinuity such that the Lorentz force of particle $2$ could be computed, however, with a jump in acceleration.

All in all, the bad trajectories and the bad initial conditions lie dense in the good ones. Or in other words, almost all initial values will not allow for global smooth solutions \citep[for the mathematical details see][pp.\ 8--9]{Deckert:2016aa}.

\newpage
\subsection{How Should Initial Fields Look Like?}
\label{subsec:initial}

In order to obtain physical and mathematically well-defined Maxwell-Lorentz solutions, one needs to get rid of shocks in many-particle systems. 
This amounts to finding \emph{compatibility conditions} between the actual solution trajectories $t \mapsto \vec{q}_{i,t}$ 
and the initial fields $\vv f_{i,0}$, where these have been parameterized by the
auxiliary trajectories $t \mapsto \tilde{\vv q}_{i,t}$ (see equation \eqref{eq:f_0}).

The first thing we need to require is that the actual and auxiliary trajectory need to match at $t=0$, that is,  $\vec{q}_{i,0}=\tilde{\vv q}_{i,0}$. Otherwise the Maxwell's constraints \eqref{eq:maxwell_constraints_particle} would be violated.

If the initial velocities do not match, we get shock fronts in form of $\delta$-distributions. If the velocities match but not the accelerations, we will have discontinuous jumps on the light-cone. In order to have smooth solutions, all higher-order derivatives need to match as well. Summarizing this in mathematical language we get:
\begin{itemize}
\item[-] By Maxwell's constraints:
$\tilde{\vv q}_{i,0}=\vec{q}_{i,0}$,
\item[-]
To rule out singularities on the light-cone: $\underset{t \nearrow 0}{\lim} \tilde{\vv p}_{i,t} =\underset{{t \searrow 0}}{\lim} \vv p_{i,t}$,
\item[-]
To rule out discontinuities on the light-cone: $\underset{t \nearrow 0}{\lim} \frac{d^2}{dt^2}\tilde{\vv q}_{i,t} = \underset{t \searrow 0}{\lim}\frac{d^2}{dt^2}\vv q_{i,t}$,
\item[-]
To secure smoothness on the light-cone: $\forall k > 2: \underset{t \nearrow 0}{\lim} \frac{d^k}{dt^k}\tilde{\vv q}_{i,t} = \underset{t \searrow 0}{\lim}\frac{d^k}{dt^k}\vv q_{i,t}$,
\end{itemize}
where the upward and downward arrows represent one-sided limits approaching from the negative or the positive numbers respectively.
 
Let's assume we were given a particle, an initial field of the form \eqref{eq:f_0}, and only the first compatibility condition, that the actual and auxiliary trajectories match at $t=0$. What happens according to the Maxwell equations is that inside the light-cone of the particle, the field is determined just by the actual trajectory, while the field outside the light-cone is determined just by the auxiliary trajectory. 
The Maxwell equations treat the initial fields (i.e. the auxiliary history) and the particle trajectories independently, and so no wonder that there will be odd behavior at the points, where both time-evolved fields meet, namely on the light-cone. The more the auxiliary trajectory matches the actual trajectory at $t=0$ the more regularized is the behavior on the light-cone. 

Now assume that we have a good initial value $(\vv q_{i,0},\vv p_{i,0},\vv f_{i,0})$ and the initial field is generated by an auxiliary trajectory that ran smoothly into the actual solution trajectory $t\mapsto \vv q_{i,t}, t\geq 0$. 
We would like to know how the real trajectory has looked like before $t=0$. This is still unknown because we started from the initial conditions of the particle at $t=0$. In principle, the actual trajectory may be propagated toward the past in infinitely many ways, while still meeting the local compatibility conditions at $t=0$. But only the auxiliary trajectory generates the initially given field at $t=0$. Therefore, the real trajectory in the past should be equal to the auxiliary trajectory, which we initially introduced solely to parametrize the fields. In this case, the initial fields simply look like this:
\begin{equation}
\label{eq:f_0_new}
\vv f_{i,0} = \lambda \vv f^-_{i,0}+(1-\lambda) \vv f^+_{i,0}+\vv f_{i,0}^{\text{free}},
\end{equation}
where instead of some auxiliary trajectory we plug in the real ones into the Li\'enard--Wiechert fields $\vv f^-_{i,0}$ and $\vv f^+_{i,0}$. 
In order to match the notation of \citet{Deckert:2016aa}, an initial free field $\vv f_{i,0}^{\text{free}}$ is assigned to each single particle; the entire free field is then the sum $\vv f_t^{\text{free}} = \sum_{j=1}^N \vv f_{j,t}^{\text{free}}$ for all $t$.
Moreover, one should note that the parameterization of the initial fields is mathematically not unique, because $\vv f^- - \vv f^+$, for instance, is always a solution of the free Maxwell equations. From an ontological point of view it is, however, relevant what portions of advanced and retarded interactions one chooses.

With equation \eqref{eq:f_0_new} the degrees of freedom of the initial fields reduce to the actual trajectories of particles $t \mapsto \vv q_{i,t}$, the free fields $\vv f_{i,0}^{\text{free}}$, and $\lambda \in [0,1]$, the proportion of advanced and retarded fields. 
And these initial fields cure the problem of shock fronts.

\subsection{Classical Electrodynamics without Shock-Fronts}
\label{subsec:consistent-dynamics}
Since the Liénard--Wiechert fields $\vv f^-_t$ and $\vv f^+_t$ are explicitly known for given charge trajectories (see \eqref{eq:lw_e_field}) and also the evolution of free fields is unique and known, the Maxwell equations become redundant. From equation \eqref{eq:f_0_new} the time-evolved fields due to charge $i$ can then be given explicitly for any time $t$, which is 
\begin{equation}
\label{eq:f_t_new}
\vv f_{i,t} = \lambda \vv f^-_t[\vv q_i, \vv p_i]+(1-\lambda) \vv f^+_t[\vv q_i, \vv p_i]+\vv f_{i,t}^{\text{free}}.
\end{equation}
Thus, one could axiomatically introduce the Liénard--Wiechert fields and plug them into the Lorentz force law:
\begin{equation}
\label{eq:LE_new}
\frac{d}{dt}\begin{pmatrix}
\vv q_{i,t}\\ \vv p_{i,t}
\end{pmatrix} = \begin{pmatrix}
\vv v(\vv p_{i,t})\\ \sum_{j \neq i} \vv E_{j,t}(\vv q_{i,t})+ \vv v(\vv p_{i,t})\wedge \vv B_{j,t}(\vv q_{i,t})
\end{pmatrix},
\end{equation}
where $\vv f_{i,t}=(\vv E_{i,t},\vv B_{i,t})$ is determined by \eqref{eq:f_t_new}. This is a reformulation of the Maxwell-Lorentz theory for point charges without self-interaction where the phenomenon of shock fronts is cured.

There are three questions left:
\begin{enumerate}
\item What happens to the initial value formulation? 
\item Can we constrain $\lambda$, the portion of advanced and retarded fields?
\item How should we choose the initial free fields $\vv f_{i,0}^{\text{free}}$? 
\end{enumerate}
We deal with these questions in the next section.

\section{Discussion}

 \subsection{From Initial Value Problems to Delay Problems}\label{subsec:ivp-delay}
 
As argued in Section \ref{sec:problem-iv}, in order to define meaningful initial fields, the whole history of charges needs to be known. 
The resulting dynamics becomes \eqref{eq:LE_new}, in which Maxwell's equations are now redundant. 
In the reformulation, the force acting on each charge, depends on the charge trajectory of all other charges at retarded and/or advanced times (depending on the choice of $\lambda$). In order to solve the system \eqref{eq:LE_new} for a given initial configuration $(\vv q_{i,0},\vv p_{i,0})_{1 \leq i \leq N}$, one would need the field value at time $0$ which is a function of position, momentum, and accelerations at the retarded and/or advanced times, which are not known. 
So this is no longer an initial value problem!  
In fact, Maxwell's equation and the Lorentz force law, a set of ordinary differential equations and partial differential equations, has turned into a system of ordinary \emph{delay differential equations}. And these delays are responsible for the system not being solvable in the conventional sense.

\begin{figure}[ht]
    \begin{center}
       \includegraphics[width=0.6\linewidth]{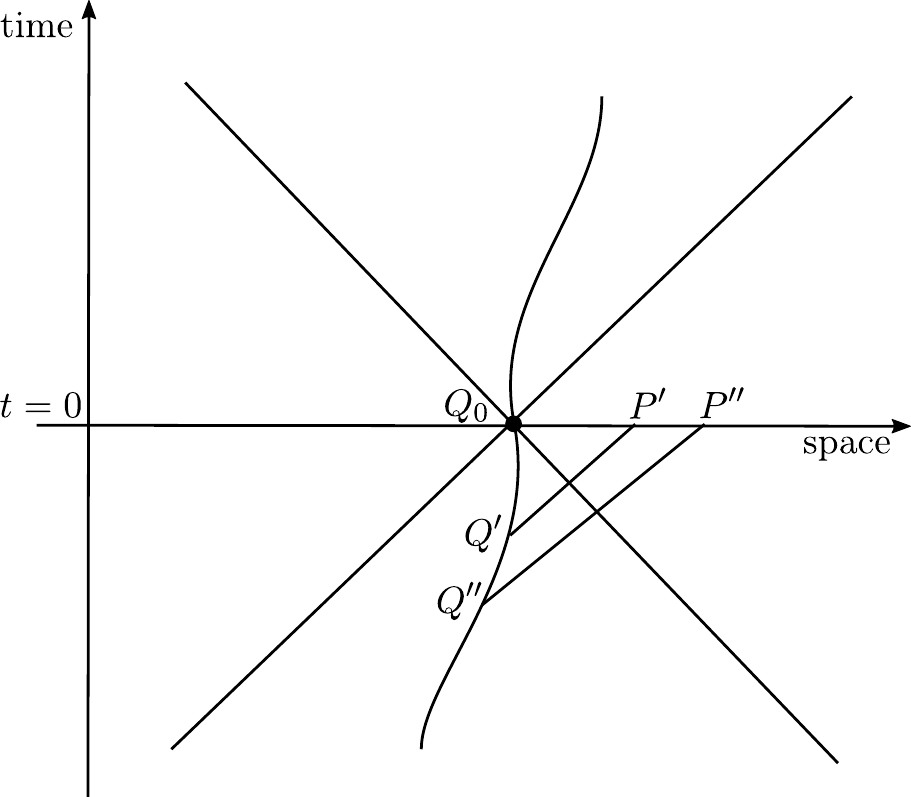} 
    \end{center}
    \caption{Image illustrating the quote by Rohrlich.}
    \label{fig:rohrlich-light-cone}
\end{figure}

It seems that some physicists have recognized that the data on a Cauchy slice needs to be constrained, although not taking this issue so seriously to doubt the role of fields:

\begin{quotation}
If one wants to specify a Cauchy problem at $t = 0$ together with the current for $t>0$, the problem will separate into two problems: (a) the Cauchy problem with Cauchy data on $t = 0$; this will determine the fields for $t > 0$ \emph{outside} the light-cone whose vertex is $Q_0$ [Fig.\ \ref{fig:rohrlich-light-cone}]; (b) the retarded field problem due to the current at $t = 0$; this will determine the fields \emph{inside} and on the future light-cone with vertex at $Q_0$. The Cauchy data for problem (a), however, are not known and must be found by solving a problem of type (b) for $t < 0$. Thus one simply has a retarded field problem [type (b)] \emph{for all space-time}. It is very essential to realize that the finite propagation velocity of the field forces one into a problem posed for \emph{all} space-time which would be very difficult (and physically awkward) to specify as (partially) a Cauchy problem. \citep[][p.\ 78]{Rohrlich:2007aa}
\end{quotation}

Rohrlich is aware that one needs to tackle a delay problem instead of a Cauchy problem. The points $P^{\prime}$ and $P^{\prime\prime}$ in Fig.\ \ref{fig:rohrlich-light-cone} on the Cauchy slice $t=0$ are determined by the behavior of particles in the past, namely, by $Q^{\prime}$ and $Q^{\prime\prime}$. This would lead to ``difficult and (physically awkward)'' delay differential equations.

So how does one solve this type of equations and how do initial data look like? 
Consider, for instance, a first-order delay differential equation of the form
$\dot{x}_t = f(t, x_t, x_{t-1})$. Then the solution at time $t$ depends on the solutions at the delayed time $t-1$. However, if one fixes the function $t\mapsto x_t$ on the interval $[0,1]$ the solution can be obtained by means of techniques from ordinary differential equations on the interval $[1,2]$. And with this new piece of function one can proceed. The sketched procedure is called \emph{the method of steps}.

\begin{figure}[ht]
    \begin{center}
       \includegraphics[width=0.7\linewidth]{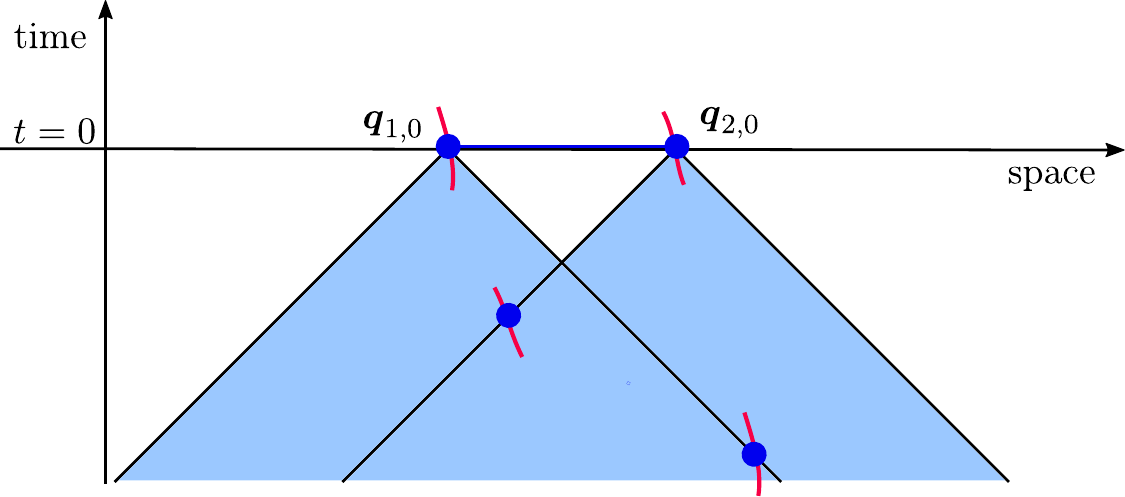} 
    \end{center}
    \caption{Time intervals where initial trajectory pieces need to be defined as well as the places where the compatibility conditions need to be met. The lower red trajectory piece of charge 1 determines the Lorentz force of the upper red piece from charge $2$ and vice versa, and, these have to match.}
    \label{fig:figureIVd}
\end{figure}

Now, assuming there are only retarded interactions one could reproduce the technique from the example for the system \eqref{eq:LE_new}. Taking into account also advanced interactions the procedure would become highly opaque though. 
In order to solve \eqref{eq:LE_new}, as initial data at least pieces of trajectories corresponding to single particles that go back to the retarded times of the initial particle positions are needed, that is, for two charges one needs to specify the time where the backward light-cone of charge 1 hits the past trajectory of charge 2 and vice versa (see Fig.\ \ref{fig:figureIVd}). 

But these initial trajectory pieces cannot be freely chosen! In order to prevent shock fronts the actual trajectory building up from time $t=0$ needs to match the past trajectory generating the fields at $t=0$. 
As the solutions is not yet known, one needs to translate the condition
\begin{equation}
 \lim_{t \nearrow 0} \frac{d^k}{dt^k} \begin{pmatrix}
\vv q_{i,t} \\ \vv p_{i,t}
\end{pmatrix} 
=
 \lim_{t \searrow 0} \frac{d^k}{dt^k} \begin{pmatrix}
\vv q_{i,t} \\ \vv p_{i,t}
\end{pmatrix}
\end{equation}
from Section \ref{subsec:initial} into a a condition that is formulated in terms of the histories only, and thanks to the Lorentz equations such an equivalent conditions is given by 
 \begin{equation}
\label{eq:I4}
\lim_{t \nearrow 0} \frac{d^k}{dt^k} \begin{pmatrix}
\vv q_{i,t} \\ \vv p_{i,t}
\end{pmatrix}
=
\lim_{t \searrow 0} \frac{d^{k-1}}{dt^{k-1}} \begin{pmatrix}
\vv v( \vv p_{i,t}) \\   \sum_{j \neq i} \vv E^-_{j,t}(\vv q_{i,t})+\vv E^{\text{free}}_{j,t}(\vv q_{i,t}) + \vv v(\vv p_{i,t}) \wedge (\vv B^-_{j,t}(\vv q_{i,t})+\vv B^{\text{free}}_{j,t}(\vv q_{i,t})
\end{pmatrix}
\end{equation}
(see Fig.\ \ref{fig:figureIVd}).

\begin{figure}[ht]
    \begin{center}
       \includegraphics[width=0.6\linewidth]{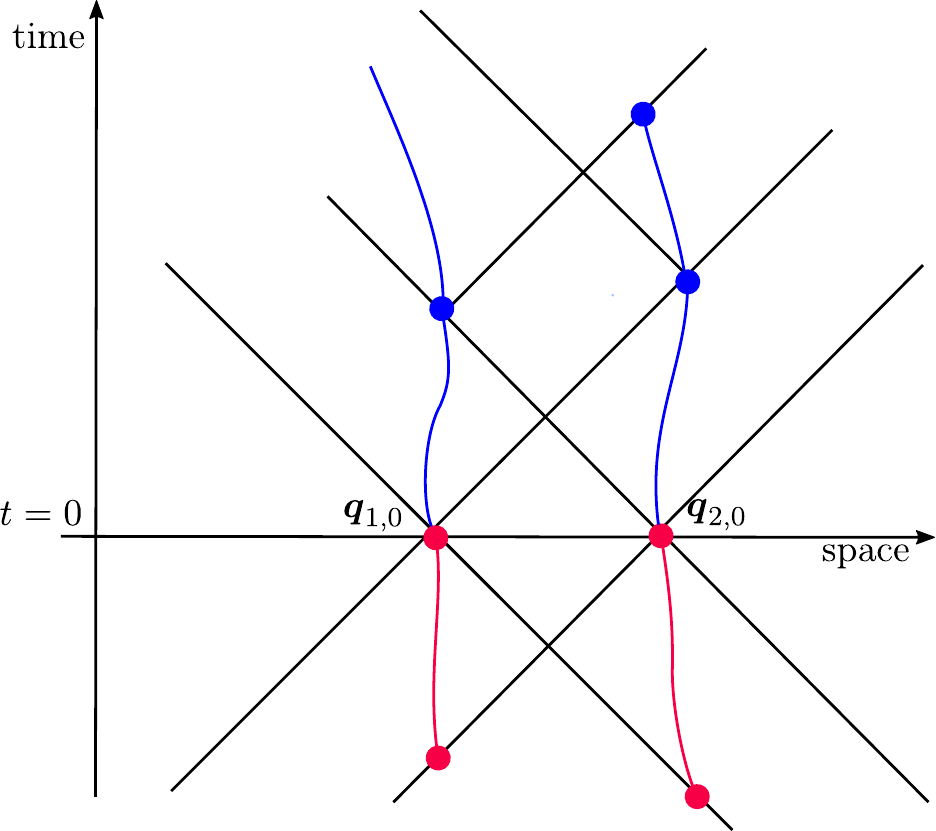} 
    \end{center}
    \caption{Illustration of the method of steps for two particles. Initial data depicted colored red; iterated solution trajectories colored blue.}
    \label{fig:steps}
\end{figure}

As the compatibility condition between the solution and the history is translated by \eqref{eq:I4} to a condition on the history only  
one can define valid initial trajectory pieces (see the red parts of the trajectories in Fig.\ \ref{fig:steps}). 
Once the red trajectory pieces in Fig.\ref{fig:steps} are given, these allow us to propagate solutions to the delay system up to the times where the single trajectories cross the first light-cone of the other initial positions. Beyond that time the initial data doesn't provide any more input for the force law. However, with the trajectory pieces obtained in the first step one can propagate one step further into the future and so on. One thing may still happen:  If particles come too close to each other, the factor $\frac{1}{|\vv x-\vv q^-|^2}$ from the Li\'enard-Wiechert fields may blow up and prevents the computation of further trajectory pieces.

Nevertheless, the method of steps has its shortcomings, as it provides Maxwell-Lorentz solutions on the half axis only, but not globally! The initial trajectory pieces work in the sense of providing solutions and avoiding shocks, but these initial pieces themselves may not be solutions of the delay system. This becomes clear, when propagating the future solution backwards. Then, the obtained past trajectories will almost surely not match the initial trajectory pieces! 

Could this be a starting point for global solutions? Once the future solutions are propagated into the past, the obtained past solutions could serve as new initial data for future propagation. The new future solutions have to be propagated back again, and the past trajectories forth again. If this kind of iteration converges at some point, i.e. the obtained solutions do not differ anymore from the ones in the previous step, one would indeed end up with global solutions to the delay problem. 

\subsection{Are Advanced Fields Needed?}
\label{subsec:degrees-of-freedom}

The first thought might be that there is a preferred direction of interaction from the past to the future, and interactions coming from the future contradict our experience.
So we may set $\lambda$ equal to 1. For $\vv f_{i,0}^{\text{free}} =0$ and $\lambda = 1$ the system is equivalent to the Synge equations---see also the retarded theory proposed by \citet{Ritz:1908aa} and the discussion in \citet{Frisch:2000aa}. This theory, however, doesn't include radiation damping, the experimentally verified phenomenon that charged particles are harder to accelerate than uncharged ones, because charged particles radiate when accelerated. Radiation damping in this theory may be dealt with in two ways: either one inserts self-fields, but they are ill-defined as we have seen, or radiation damping is caused by the fields of other charges in the past, but then these retarded fields won't reach the particle in time.\footnote{An anonymous referee mentioned the work of \citet{Gralla:2009aa} who claim to have derived a point-particle model that evades the self-interaction problem without evoking advanced solutions. In a nutshell, they start with a continuous charge distribution and take a point-particle limit by taking the charge and the mass to go to zero, while keeping the ratio of the mass and charge fixed. In our opinion, their work reveals an interesting mathematical result, but it leads to an unphysical model, where trajectories exist but no particles travel on them. The authors (pp.\ 2--3)  explicitly mention the disappearance of particles, ``[\dots] we consider a modified point particle limit, wherein not only the size of the body goes to zero, but its charge and mass also go to zero. More precisely, we will consider a limit where, asymptotically, only the overall scale of the body changes, so, in particular, all quantities scale by their naive dimension. In this limit, the body itself completely disappears, and its electromagnetic self-energy goes to zero.'' Beware the last sentence: there are no physical objects but in any case their self energy is no longer infinite. We doubt that this is a proper physical solution of the self-interaction problem.}

So its seems that the sole option that is left (without changing the Maxwell equations or inflating particles) is to include the effects of charges in the future, i.e., setting $\lambda \neq 1$. 
We briefly sketch an argument inspired by \citet[][p.\ 169]{Wheeler:1945aa}, to demonstrate how advanced fields may account for radiation damping.

Say charge $i$ feels an acceleration at time $t$, and thus, creates a disturbance $\vv f_{i,t}^-$, that sets in motion the $N-1$ other charges at the advanced times, which in return, act back on charge $i$ through advanced fields. 

Moreover, for simplicity, we assume that the $N-1$ other charges are sufficiently randomized, namely, such that at the same time $t$ 
\begin{equation}
\label{eq:19}
\sum_{j \neq i}\vv f_{j,t}^-(\vv q_{i,t}) = 0.
\end{equation}

The idea is to write the measured disturbance $\vv f_{i,t}^-$ of charge $i$ as a sum of the actual generated field of charge $i$ and the back reaction of the other $N-1$ charges:
\begin{equation}
\vv f_{i,t}^-
= 
\overbrace{\lambda\vv f_{i,t}^-+(1-\lambda)\vv f_{i,t}^+}^{\text{fields generated by charge }i}
+
\overbrace{
\underbrace{\lambda\sum_{j\neq i}\vv f_{j,t}^-}_{=0, \text{ by \eqref{eq:19}}} 
+ 
(1-\lambda) \sum_{j \neq i}\vv f_{j,t}^+}^{\text{fields generated by the other $N-1$ charges}}.
\end{equation}
And therefore, in the vicinity of $\vv q_{i,t}$, the following relation holds:
\begin{equation}
(1-\lambda)(\vv f_{i,t}^--\vv f_{i,t}^+) =
(1-\lambda)\sum_{j \neq i}\vv f^+_{j,t} ,
\end{equation}
where, according to \citet{Dirac:1938aa},
 $\vv f_{i,t}^--\vv f_{i,t}^+$ is the radiation friction felt by charge $i$ at time $t$ and therefore by means of the Lorentz equations \eqref{eq:lorentz} the effective force on charge $i$ goes like 

\begin{equation}
\sum_{j \neq i} \left(\lambda \vv f_{j,t}^-(\vv q_{i,t})+(1-\lambda) \vv f_{j,t}^+(\vv q_{i,t}) \right)
=
(1- \lambda)(\vv f_{i,t}^--\vv f_{i,t}^+)(\vv q_{i,t}) .
\end{equation}

And thus, taking into account advanced fields (i.e. $\lambda < 1$) we have an estimation that we can explain radiation damping. 

But what is a good choice of $\lambda$? $\lambda=1/2$ is distinguished by two features: first, it ensures the energy loss predicted by the experimentally verified Larmor formula \citep[][p.\ 612]{Bauer:2014aa}, and, second, it leads to a time-reversal-invariant theory governed by an action principle by means of the Fokker--Tetrode--Schwarzschild Lagrangian. As we are investigating a fundamental particle theory, time-reversal-invariance and energy--momentum conservation (derived from Noether's theorem) may be desirable.

For $\vv f_{i,0}^{\text{free}} =0$ and $\lambda=1/2$ we indeed end up with the Wheeler--Feynman theory, a theory which is solely about moving charges (see Appendix \ref{sec:wheeler-feynman} for a brief introduction and our take on the notorious absorber condition). So, starting with a particle-field view in electrodynamics, by a series of arguments one remarkably reaches an action-at-a distance theory \citep[for critical remarks concerning time-symmetric electrodynamics see][Ch.\ 6]{Frisch:2005fr}. And this was derived completely independently from how we  treat the self-interaction problem. Making the specific choice of $\lambda$, however, is not our focus here and does not affect the point we want to make.

\subsection{Do Free Fields Exist?}

Assume the proportion of advanced and retarded interactions is fixed. Then, the free field remains the sole free parameter. When applying electrodynamics to subsystems, the free field plays an imminent role in fixing boundary conditions. This field is free in the sense that it is not generated by charges \emph{inside} the subsystem; rather, it is coming from outside. This field was actually generated by other particles somewhere in the universe and then has been propagating towards the subsystem. There is nothing to say against this use of free fields. But this would mean that free fields are not degrees of freedom of the theory as they are completely determined by the motion of particles.

Genuine free fields, on the other hand, are generated by no particles at all. These are fields set in the initial conditions of the universe, and they only need to fulfill the homogeneous Maxwell equations. If we imagine God creating the universe, he had to set the the initial conditions of particles and the fields in such a away to avoid shock fronts. It would be possible that God created particles and some initial field without paying further attention to their initial conditions. This would almost surely amount to shock fronts. These shock fronts, however, could be compensated if God  created in addition a free field and had decided to meticulously choose very special initial conditions of this field. In doing so,  God could nullify shock fronts. This option is logically possible but would be based on a fine-tuned initial free field. But we know that good scientific explanations should be free of such conspiracies.

If the free field existed it would ripple once through the universe and disappear into the void never returning to make its march back again. So if you happen to see your TV screen flickering next time, would you bet that it might have been a free field crossing your apartment? We have no experimental evidence whatsoever that genuine free fields exist. How can we? They would be indistinguishable from fields generated by particles. 

Moreover, if these fields existed their energy would be negligible. Let's say that the free field was created at some point far back in history, say at time minus infinity. If at that time the field has been finite, say of order $\norm{\vv F^{\text{free}}_{-\infty}}_{L^2} \leq O(\frac{1}{\norm{\vv x}^\epsilon})$,
then it would vanish until the present time $t=t_0$ because it will spread. In particular,  the energy of the field would become negligible: If the energy due to the free field was finite at the beginning of the universe, that is $\int dx (\vv E^{\text{free}}_{-\infty}(\vv x))^2+(\vv B^{\text{free}}_{-\infty}(\vv x))^2 < \infty$, then the free field vanishes until $t=t_0$. One could imagine converging free fields coming from the future, which indeed would get stronger (thanks to an anonymous referee for making this point). Incoming free fields are theoretically possible but never found in experiments. Furthermore, they don't play any explanatory role apart from their possible theoretical existence, unlike generated advanced fields which would account for radiation friction (see section \ref{subsec:degrees-of-freedom}). 

\citet[][Sec.\ 2.3]{Earman:2011aa} pointed out that every solution in terms of only the retarded fields can be expressed as a sum of an advanced solution and a free field. This is supposed to show that free fields cannot be ignored.\footnote{We thank an anonymous referee for mentioning this argument.} Of course, it is a mathematical fact that the general solution of the Maxwell equations does not have a unique mathematical representation. But to switch between a purely retarded representation and a representation including an advanced and a free field would amount to change $\lambda$, although we have fixed $\lambda$ before as part of the theory. Second, Earman repeatedly emphasizes to let ``the equations speak for themselves'' in order to find the proper interpretation of the formalism. We don't see, however, how this can be done; the mathematics doesn't carry with it the proper ontology. In particular, a change of representations with a free field cannot reify a free field, nor does it commit us to a free field in the initial conditions of the universe.
\citet[][Sec.\ 6]{Lazarovici:2016aa} reacts to Earman's argument that the only mathematically distinguished representation of the free field would be a field that is everywhere zero. Such a requirement can be achieved by choosing appropriate boundary conditions.

So if genuine free fields do not exist or are negligible, the entire field turns into the following convex combination of retarded and advanced Li\'enard-Wiechert fields:
\begin{equation}
\label{eq:f_t_new_new}
\vv f_{i,t} = \lambda \vv f^-_t[\vv q_i, \vv p_i]+(1-\lambda) \vv f^+_t[\vv q_i, \vv p_i].
\end{equation}

\subsection{Do Generated Fields Exist?}

Should we we be realist with respect to the fields $\vv f_{i,t}$? The most famous argument for the existence of electromagnetic fields says that since these fields restore energy--momentum conservation, which would not hold if we just consider the energy and momenta of particles, they are real. The energy and momentum for the electromagnetic field are defined from the \emph{Poynting theorem} \citep[e.g.,][section 6.7]{Jackson:1999aa}, which states that the total energies and momenta of particles and fields are conserved. 

In mathematical notation, the Poynting theorem is:
\begin{align}
\dot{E}_{m}+\dot{E}_{f}&=\oint\limits_S\vec{S}\cdot\mathrm{d}\vec{A} , \nonumber
\\
\dot{P}_{m}+\dot{P}_{f}&=\oint\limits_S\vec{T}\cdot\mathrm{d}\vec{A}.\nonumber
\end{align}
Here, the dot stands for the derivative with respect to time. $E_{m}$ and $P_{m}$ are the mechanical energy and the mechanical momentum of particles within a volume $V$. $E_{f}$ and $P_{f}$ are the field energy and the field momentum within $V$. The right side of the equations symbolizes the energy and momentum crossing the surface $S$---$\vec{S}$ is the energy flux density (the \emph{Poynting vector}), and $\vec{T}$ is the momentum flux density (the \emph{Maxwell stress tensor}). Poynting's theorem then states: energy and momentum change inside a volume $V$ if and only if energy and momentum cross the surface $S$ of $V$.

Conservation laws are physically significant because they facilitate to solve the equations of motion. This is in our opinion the main advantage of these laws. To dub conservation laws as laws in the first place is actually a misnomer, because they are not laws of nature; rather, they are derived from the basic laws, the Maxwell equations and the Lorentz force law. Especially since special relativity, there is an intuition behind energy as a kind of stuff that flows between objects and that is in some way equivalent to mass or matter. We don't share this intuition \citep[see also][]{Adler:1987aa}. Energy and momentum are indeed frame-dependent quantities in special relativity, as shown by the relativistic energy--momentum formula, which is nothing but the length of the momentum $4$-vector. And therefore we do not agree that energy--momentum conservation reifies fields \citep[for an extensive discussion, see][Ch.\ 5 and 8]{Lange:2002ys}.

By Ockham's razor we are to be as sparse in our ontology as possible \citep[see][for an up-to-date monograph on Ockham's razor]{Sober:2015aa}. So just by naively applying the razor, we may argue against the physical existence of the fields $\vv f_{i,t}$. But Tim Maudlin has warned us to be cautious with applying the razor:  

\begin{quotation}
\emph{Entia non sunt multiplicanda praeter necessitatem} [entities should not be multiplied beyond necessity], and
the availability of a reduction obviates any necessity. Surely we should be
seeking the slenderest basis on which to erect our ontology.

But it is not clear that the Razor can withstand much critical scrutiny. If by
\emph{necessitas} one means \emph{logical} necessity, then the Razor will land us in solipsism.
But if one means something milder---entities ought not to be multiplied
\emph{without good reason}---then the principle becomes a harmless bromide: nor
should one’s ontology be \emph{reduced} without good reason. \citeyearpar[][pp.\ 3]{Maudlin:2007aa}
\end{quotation}
Ockham's razor doesn't say that a sparse ontology is always to be preferred over a not-so-sparse ontology. We need to have good reasons to cut off the ontology. We think that a good reason is that fields are not further degrees of freedom beyond the degrees of freedom of particles. Fields appear as mere book-keepers for the history of particles. This argument together with the razor would be sufficient in our opinion to dismiss fields from the ontology.

On the other hand, there is also a good reason \emph{to keep} the fields $\vv f_{i,t}$ in the ontology, namely, because fields are said to restore \emph{locality} \citep[see][Ch.\ 4]{Lange:2002ys}. In this case, locality is to be contrasted with action-at-a-distance, and the merit of fields is that they mediate the action between particles in space and time. With fields it wouldn't seem mysterious or miraculous why one particle can affect another one at a distance and at a later time. For this argument one introduces fields in the sense of the relation \eqref{eq:f_t_new_new} without referring to the Maxwell equations. Although fields can be reduced to the motion of particles, one could still reify them because they restore locality. 

But the lack of fields doesn't mean that there is nothing that is able to account for mediating the action; it rather means that \emph{fields} are not fulfilling this job.\footnote{We thank an anonymous referee for making helpful suggestions in this passage.} If there is no physical mechanism, that is, something like fields introduced by the physical theory to mediate, there may be \emph{metaphysical} means. And indeed there are those metaphysical tools that demystify or ground action-at-a-distance. In order to provide a real connection between particles one may introduce a dynamical ontic structure, as proposed by \citet[][]{Esfeld:2009aa}. This ontic structure would be a real structure in space and time constraining the motion of particles. Even if the physical theory is void of a physical mediator, like a field, an action-at-a-distance theory may be embedded in a metaphysics of ontic structures. These structures provide for the modal connections among particles. This strategy has been worked out in quantum mechanics \citep[see, for instance,][for an overview]{Lam:2015aa}. A detailed application of this idea to action-at-a-distance theories of electromagnetism is still missing. We guess that it may be possible to introduce dynamical ontic structures on Minkowski space as relations between events on the light-cones. 

It is also possible to reduce the dynamical relations to laws themselves and adhere to a primitivism about laws as proposed by \citet{Maudlin:2007ah}. In this case, there won't be any dynamical relations in space and time mediating between particles; instead, the laws themselves would have the power to guide particles and to generate the time evolution. For doing so, one needs to grant laws this metaphysical weight, otherwise the modal connection via laws would be mysterious. This is another viable option that is worth pursuing. 

A Humean, however, cannot use these metaphysical tools. Even in a Humean field theory, there is no modal connection between the motion of particles and the behavior of fields ``generated'' by them, and there is no modal connection between the motion of one particle and the motion of another particle. Moreover, there is even no modal connection between the motion of a particle and the field value at the location of the particle. Pace \citet[][Ch.\ 6]{Albert:2015aa}, the geometrical relationship between particles and fields---that particles move in the direction the field is pointing---is not what we would expect in a Humean world. So we see that the Humean ontology is much weaker than a naive interpretation of an action-at-a-distance theory. A naive interpretation would go along these lines: somehow particles are modally connected but the physical theory doesn't reveal this connection. Humeans are not convinced of locality in the sense of having something in space and time that mediates the influences from one particle to the other. They are rather motivated by another reading of locality, which is more appropriately dubbed \emph{separability}: all physical phenomena can be reduced to local matters of particular facts. Therefore, there doesn't seem to be a reason to have fields in a Humean ontology; they are just good means for an efficient description of the going-ons in the mosaic \citep[see also][]{Vassallo:2016aa}.

All in all, we face a physical fact: the electromagnetic field is not a further degree of freedom. Should we now pair this result with Ockham's razor or with locality? By applying the razor, the Maxwell--Lorentz theory would be indeed an action-at-a-distance theory disguised as a field theory. Having uncovered the true mathematical framework of this theory, it cannot be distinguished from an action-at-a-distance theory. In particular, if the advanced and retarded fields comprise each one-half of the entire field, the theory reduces to the Wheeler--Feynman theory.  But if locality outweighs Ockham's razor, the Maxwell--Lorentz theory can be interpreted as a field theory, where the field is mathematically, but not ontologically, reduced to the motion of particles. Although the mathematical structure would still suggest an action-at-a-distance theory, one may construe the field values as referring to something real in space-time.

In the end, we think that regarding Maxwell--Lorentz a proper field theory would be very artificial after our analysis of the initial value problem. Locality seems to be a common sense requirement for science. In particular, one would not naively understand how we can send and receive radio signals if there weren't electromagnetic fields. In our opinion, however, there are satisfying tools, although metaphysical ones, for explaining the modal connections in our world.

\subsection{Inflated Particles to the Rescue?}

Point-like particles are often blamed for being unphysical and for being the actual source of singularities. 
The Abraham model where particles are modeled as balls with non-zero diameter cures the problem of undefined self-interactions. In fact, this model is mathematically well understood because the singularities at the charge positions are smeared out \citep[see][]{KS00,BDD10}.
Also the singularities on the light cones are successfully cured in this model. Nevertheless, the singularities in the fields along the light cones are just smeared out and persist in form of physically questionable increases in the field and respectively in increases in acceleration for particles that cross the corresponding light cone. 
This phenomenon can be taken from a quantitative example in \citep{Deckert:2016aa} which demonstrates that even in the Abraham model bad initial values lead to a flank in radiation power which should be measurable but is not observed in experiments. Moreover, they show, that one flank results in a whole network of flanks (see again Figure \ref{fig:network}).
In the end, the choice of initial values remains a problem to deal with, whether or not the charges are extended distributions or not. Though mathematically sound in the case of smeared out charges, initial values which do not comply with the Lorentz force lead to physically questionable behavior.

\section{Conclusion}

We have argued that the Maxwell equations are redundant once the Liénard--Wiechert fields are known. Instead, the entire theory of classical electromagnetism is contained in equations \eqref{eq:f_t_new} and \eqref{eq:LE_new}. In the end, one needs to solve delay differential equations, whose solution theory is still to be explored. In particular, we have illustrated the following two features of the electromagnetic field:
\begin{enumerate}
\item Generic initial values lead to ill-defined trajectories.
\item The field is not a degree of freedom beyond the degrees of freedom of particles. It rather encodes the motion of particles, and so it's functional role is  to be a book-keeper of the behavior of charged particles.
\end{enumerate}
We think that these features should be an incentive to proceed with developing field-free formulations of classical electrodynamics, like the Wheeler--Feynman theory (see Appendix \ref{sec:wheeler-feynman}). In such a theory, fields may appear as tools for dealing with subsystems, but they would be no longer independent from particles. 

We also appreciate further development of alternative field theories, like the Born--Infeld theory and the Bopp--Podolsky theory. That fields are not further degrees of freedom may also affect these theories, and further research has to clarify whether these field theories also suffer from the same problem of initial values as the Maxwell--Lorentz theory \citep[for results in the Bopp--Podolsky theory, see][]{Kiessling:2017aa}.

Hence, it is open to philosophical debate which framework would be the best for classical electrodynamics: a field theory with no field degrees of freedom or a proper action-at-a-distance theory. Following the field theory approach, where there are no field degrees of freedom, would challenge our understanding of what field theories really are and how they differ from action-at-a-distance theories. As we saw, the equations of motion of the Maxwell--Lorentz theory reduce to the very same delay differential equations one encounters in the Wheeler--Feynman theory (by fixing $\lambda = \frac{1}{2}$).

So what would distinguish a field theory from an action-at-a-distance theory?
Even though a field theory and an action-at-a-distance theory may share the same dynamical equations, they may not be ontologically equivalent. Indeed, mathematical equivalence between two theories doesn't logically entail ontological equivalence. Further arguments are needed why one ontology is to be preferred from the other. We  briefly discussed energy--momentum conservation, Ockham's razor, and locality. Of all these arguments, locality is in our opinion the strongest to reify fields. On the other hand, there are metaphysical tools to establish modal connections for action-at-a-distance theories. Then direct interactions would not be so miraculous. 

\section*{Acknowledgements}

We wish to thank Michael Esfeld and Dustin Lazarovici for helpful comments on previous drafts of this paper. Special thanks go to Dirk-André Deckert and Michael Kiessling for many hours of fruitful discussions. We also thank three anonymous referees for their meticulous reviews. Funding for this research was provided by the Elitenetzwerk Bayern (to Vera Hartenstein as part of the research group \emph{Interaction of Light and Matter}) and by the Swiss National Science Foundation (to Mario Hubert as part of the Early Postdoc.Mobility Fellowship, grant no.\ 174745).  

\appendix

\section{Alternative Field Theories with Point Charges}
\label{sec:appendix}

We briefly mention the two alternatives to a field theory with point charges, since they are   underrepresented in the philosophical literature.

\subsection*{The Born--Infeld Theory}

In the Maxwell--Lorentz theory, the relations between the electric field $\vec{E}$, the electric displacement field $\vec{D}$, the magnetic field $\vec{B}$, and the magnetic induction field $\vec{H}$ in vacuum are trivial, namely, $\vec{E}=\vec{D}$ and $\vec{B}=\vec{H}$. \citet{Born:1934aa} proposed to replace these \emph{constitutive relations} by non-linear relations between $\vec{E}$, $\vec{D}$, $\vec{B}$, and $\vec{H}$ in vacuum. 

The Born--Infeld self-field of a static particle is bounded 
\begin{equation*}
\vec{E}=\frac{q}{4\pi \sqrt{r_0^4+r^4}}\vec{e}_r,
\end{equation*}
with $r_0^2=\frac{q}{4\pi b}$ \citep[see][p.\ 530]{Perlick:2015aa}. The constant $b$, called \emph{Born's field strength}, is a new constant of nature in the Born-Infeld theory. We immediately see that $\left|\vec{E}\right|\rightarrow b$ if $r\rightarrow 0$; that is, the absolute value of the electric field is finite.

The dynamical case is very hard and poses many obstacles \citep{Kiessling:2012aa}. Since the Born--Infeld equations are non-linear there are no standard methods for solving them. There are no solutions for the field of a moving particle analogous to the Liénard--Wiechert fields. Even qualitative results are difficult to deliver. So no one knows whether the self-force and the self-energy are finite for dynamic systems \citep[for important results regarding the static case, see][]{Kiessling:2011aa,Kiessling:2012ab}.

\subsection*{The Bopp--Podolsky Theory}

In the 1940s, \citet{Bopp:1940aa}, \citet{Podolsky:1942aa}, and  \citet{Lande:1941aa} independently developed another field theory aimed at taming the self-interaction problem. This theory is linear but of higher order than the Maxwell equations. The Bopp--Podolsky analogue of the Coulomb field reads
\begin{equation*}
\vec{E}=\frac{q}{4\pi r^2}\left(1-\left(\frac{r}{l}+1\right)\mathrm{e}^{-\frac{r}{l}}\right)\vec{e}_r,
\end{equation*}
with $l$ being a new constant of nature. The modulus of this electrostatic field is also finite at the origin, that is, when $r\rightarrow 0$ then $\left|\vec{E}\right|\rightarrow \frac{q}{8\pi l^2}$.

Unlike the Born--Infeld theory, an analogue of the retarded Liénard--Wiechert potential has been calculated \citep[see][]{Gratus:2015aa,Perlick:2015aa}:
\begin{equation}
A(x)=\left(\int\limits^{\tau}_{-\infty}\frac{J_1(\frac{s(x,\tau^{\prime})}{l})}{l_s(x,\tau^{\prime})}\dot{z}^a(\tau^{\prime})\,\mathrm{d}\tau^{\prime}\right)\eta_{ab}\,\mathrm{d}x^b,
\end{equation}
with $J_1$ as the Bessel function of the first kind. We don't need to explain all the details of this potential, but one feature is important: the field value at $x$, lying on the future light-cone of $z(\tau)$, depends on the entire trajectory $z(t)$ from $t=-\infty$ to $t=\tau$. This is one of many ways to write down the vector potential $A$. In their calculation, \citet{Gratus:2015aa} assumed that the entire particle trajectory is known until $\tau$.

This mathematical representation of the potential does not hinder a well-posed initial-value problem in the Bopp--Podolsky theory (Michael Kiessling, private conversation). The initial fields in this theory need to be constrained similarly to the Maxwell--Lorentz theory. It turns out, however, that the shock fronts on the future light-cones in the Bopp--Podolsky theory are one order milder. Singularities in the form of $\delta$-distributions do not occur; the severest singularities are discontinuous jumps. So if one matches the initial velocity of the actual trajectory with the initial  velocity of the auxiliary velocity, one already gets continuous global fields. Furthermore, one doesn't need to know the past behavior of particles. The only singularity that may occur is when two (or more) particles collide \citep{Kiessling:2017aa}.

\section{The Wheeler--Feynman Theory}
\label{sec:wheeler-feynman}

The action-at-a-distance theory named after \citet{Wheeler:1945aa,Wheeler:1949aa} is cured of the self-interaction problem since fields as further degrees of freedom no longer exist. And so the equation of motion for a single point particle is well-defined. The price of the Wheeler--Feynman theory, however, is that advanced effects play a fundamental role. For the description of these effects, one can formally define fields, so that the entire ``field'' at some point $\vec{x}$ consists of the sum of the retarded and advanced field: $\frac{1}{2}\left(\vv f^{-}+\vv f^{+}\right)$. 

The Wheeler--Feynman theory needs to meet two challenges. First, it needs to account for radiation damping, the effect that charged particles are harder to accelerate than uncharged ones by losing energy. Second, it needs to convincingly justify why there is typically no advanced radiation to be observed in subsystems, although the fundamental laws require both retarded and advanced actions.

Both challenges can be met by a statistical analysis of the theory. Radiation damping is generated by the interaction of a particle with particles in the environment. The advanced action coming from the environment are responsible for the additional force one needs to exert to accelerate a charged particle. So radiation damping is explained as a statistical phenomenon in a many-particle system; radiation damping cannot occur in a one-particle universe according to the Wheeler--Feynman theory. Moreover, one can show that in typical situations advanced fields cannot be observed by the same statistical reasoning \citep[see][]{Bauer:2014aa}. Indeed, this argument was already given by \citet[][p.\ 162--5]{Wheeler:1945aa}, but they wanted to go a step further in postulating a meta-principle for the behavior of the environment, the infamous \emph{absorber condition}. The theory has been dismissed for the (unjustified) validity of this condition \citep[see, for instance,][Ch.\ 6]{Frisch:2005fr}. But as we sketched, the absorber condition is not needed in explaining radiation damping.

To still dub the theory without the absorber condition \emph{Wheeler--Feynman theory} would be indeed a misnomer since all what is needed to explain radiation damping had already been developed by \citet{Fokker:1929aa}, \citet{Schwarzschild:1903aa}, and \citet{Tetrode:1922aa}. Still we don't diverge from the tradition and continue to call this theory the Wheeler--Feynman theory.

\bibliographystyle{abbrvnat}
\bibliography{references}

\end{document}